\documentclass[lettersize,journal]{IEEEtran}

\usepackage{amsmath,amssymb,amsfonts}
\usepackage{algorithmic}
\usepackage{algorithm}
 \usepackage{array}
\usepackage{stackengine}
 \usepackage{textcomp}
 \usepackage{stfloats}
 \usepackage{url}
 \usepackage{verbatim}
 \usepackage{graphicx}
  \usepackage{multirow}
 \usepackage{subcaption}
 \usepackage{color}
 \usepackage{hyperref}
 \usepackage{balance}

   \def\vx{{\bf x}}
\def\vy{{\bf y}} \def\vz{{\bf z}}

\def\vY{{\bf Y}} 

\def\vSigma{{\boldsymbol\Sigma}} 
\def\vmu{{\boldsymbol \mu}}

\begin{document}

\title{An Improved Compound Gaussian Model for Bivariate Surface EMG Signals Related to Strength Training}
\author{Durgesh Kusuru, \IEEEmembership{Student Member, IEEE}, Anish C. Turlapaty \IEEEmembership{Senior Member, IEEE} and \\ Mainak Thakur  \IEEEmembership{Member, IEEE}

\thanks{
The authors are with the Bio-signal Analysis Group, Indian Institute of Information 
Technology Sri City, Chittoor, Andhra Pradesh, 517646, India. 
(e-mails: durgesh.k@iiits.in, anish.turlapaty@iiits.in and mainak.thakur@iiits.in)}
This work has been submitted to the IEEE for possible publication. Copyright may be transferred without notice, after which this version may no longer be accessible.
}



\maketitle

\begin{abstract}
Recent literature suggests that the surface electromyography (sEMG) signals have non-stationary statistical characteristics specifically due to random nature of the covariance. Thus suitability of a statistical model for sEMG signals is determined by the choice of an appropriate model for describing the covariance. The purpose of this study is to propose a  Compound-Gaussian (CG) model for multivariate sEMG signals in which latent variable of covariance is modeled as a random variable that follows an exponential model. The parameters of the model are estimated using the iterative Expectation Maximization (EM) algorithm. Further, a new dataset, electromyography analysis of human activities - database $2$ (EMAHA-DB2) is developed. Based on the model fitting analysis on the sEMG signals from EMAHA-DB2, it is found that the proposed CG model fits more closely to the empirical pdf of sEMG signals than the existing models.  The proposed model is validated by visual inspection, further validated by matching central moments  and better quantitative metrics in comparison with other models. The proposed compound model provides an improved fit to the statistical behavior of sEMG signals. Further, the estimate of rate parameter of the exponential model shows clear relation to the training weights. Finally, the average signal power estimates of the channels shows distinctive dependency on the training weights, the subject's training experience and the type of activity.  
\end{abstract}


\section{Introduction}
\subsection{Background}

{

\IEEEPARstart{S}{tatistical} models of strength of surface electromyography (sEMG) signals have many applications including a) to develop insights into sEMG signal generation from the constituent motor unit action potentials (MUAPs) that forms a basis for the sEMG signal synthesis \cite{wang2006simulation} and simulation studies \cite{stashuk1993simulation}, b) to enhance the interpretation of the sEMG signals in clinical studies such as neuromuscular disorders detection \cite{cuddon2002electrophysiology}, c) to improve performance for pattern classification of intent to control wearable exoskeleton and prostheses \cite{fleischer2006application}, d) to improve system identification models that non-invasively determine muscle force and joint torque \cite{clancy1999probability}, e) to understand interrelationships between sEMG signals and muscle groups, for example in sports activities \cite{vigotsky2018interpreting}, \cite{de1997use}, \cite{soderberg1984electromyography}, and f) to build visualization tools to support movement sciences \cite{hasanbelliu2004multi}, and muscle physiology examinations and the sports science education.
sEMG signals can be modelled as stochastic processes because each constituent motor unit firing can be considered a random event \cite{furui2019scale}. Many studies \cite{parker1977signal,hogan1980myoelectric,der1998detection} have attempted to extract their features by analyzing EMG signals, which are typically assumed to follow the Gaussian distribution. In one of the earliest experiments \cite{parker1977signal}, the Gaussian distribution was used to explain the statistical nature of EMG signals. In a similar work, Hogan et al \cite{hogan1980myoelectric}  used the Gaussian model to describe the relationship between EMG signals and the muscle force. Moreover, they assumed that EMG signals have a constant variance under constant force conditions. However, even under constant force, sEMG signals may not follow a steady Gaussian distribution \cite{milner1975relation,hunter1987estimation,clancy1999probability,bilodeau1997normality,naik2011kurtosis}. A study by Milner-red et al \cite{milner1975relation} showed that in the presence of constant-force conditions, the distribution of EMG signals collected from the bicep and first dorsal interosseous muscles underwent a sharper peak than that of the Gaussian distribution. A few simulation studies \cite{zhao2012simulation,messaoudi2017assessment}  show that the non-Gaussianity of EMG signals differs according to the level of muscle contraction, so that as the muscle contraction level increases, the distribution of EMG signals shifts towards the Gaussian.

It is well known that the Compound-Gaussian model is usually employed for modeling the heavy-tailed distributions \cite{gini2000performance,yao2003spherically,furui2021emg}. Recently, Furui et al \cite{furui2019scale} proposed a  scale-mixture model to account for the non-stationarity of sEMG signals at different muscle contraction forces. From these studies \cite{hayashi2017variance,furui2019scale}, it is evident that the variance of univariate sEMG signals is random in nature. In this study, we investigate the non-stationary models for multi-variate sEMG signals. 
In which case, the variance in univariate models is replaced by a random variable that represents the latent variable of the covariance matrix. For example, in \cite{furui2021emg}, a multivariate compound model was proposed where the a latent variable follows an inverse gamma (IG) distribution.  However, the suitability of the IG distribution was not evaluated through comparison with other possible distributions commonly used in compound Gaussian modeling. Some of the other possible models of this latent variable include Gamma, exponential and inverse Gaussian distributions \cite{wang2006maximum,furui2019scale}. Hence identification of a suitable distribution for the latent variable of the covariance that best fits the non-stationary sEMG signal characteristics is the focus of this study.

The major contributions of this study are as follows. A compound Gaussian model is proposed for the non-stationary surface EMG signals with the latent variable of the covariance following an exponential distribution. A new dataset of sEMG signals corresponding to weight training exercises under isotonic and isometric contractions is developed and named \textit{electromyography analysis of human activities - database 2} (EMAHA-DB2). The proposed model is tested on EMAHA-
DB2 and its suitability is compared against the existing models using both the qualitative and quantitative approaches. Finally, the rate parameter ($\lambda$) of the proposed model and the multi-channel signal power
are analyzed for their dependencies on different measurement conditions.

The rest of the work is organised as follows: Section-II presents the proposed model and its parameter estimation using Expectation Maximization (EM) algorithm, followed by model validation methods. Section-III describes the dataset, Section-IV presents model analysis and discussion. Finally, Section-V concludes the work. 
  
\section{Statistical Model and Problem Description}
\subsection{A Compound Gaussian Model}
A compound probabilistic model is proposed for the strength of multi-variate sEMG signals. Specifically, the multi-channel signal is modelled as product of two interacting random processes. The first component is a fast changing sEMG signal strength and the second component is a slow varying latent random variable that represents temporal fluctuations in the covariance of the observations. Thus proposed model for the multichannel sEMG observations $\vy_{n,k}$ is
\begin{equation}
     \vy_{n,k}=\vmu+\sqrt{z_{k}} \vx_{n,k}
     \label{signal_model1}
\end{equation}
Here $\vx_{n,k}$ represents the fast changing multi-variate random process within each $k$-th segment and $z_k$ is the slow changing hidden variable. Borrowing from the literature on compound Gaussian models for radar clutter \cite{richards2014fundamentals}, the variable $z_k$ will be henceforth referred as the texture.
The variations in each phase of hand activity can be attributed to the texture $z_k$ of $k$-th segment. Here $\vmu$ denotes mean vector, $T = N \times K$ is the total number of observations in each channel, $K$ denotes the number of segments in each channel and $N$ denotes the number of observations within each $k$-th segment. {The model analysis and the parameter estimation is carried out for $(K,N)=(325,40)$ and the justification of this choice is given in sec. \ref{KLD-SECTION}}. An illustration of a two channel sEMG signal relating to the compound statistical model is shown in Fig. \ref{KN_PLOT}. The probability density function (pdf) of $\vy_{n,k}$ conditioned on the texture  $z_k$ is defined as 
\begin{eqnarray}     
    p({\vy_{n,k}|z_k}) = \frac{1}{(2\pi z_{k})^{d/2}\left| \vSigma  \right |^{1/2}} \exp \bigg(-\frac{Q(\vy_{n,k})}{2 z_k} \bigg)
    \label{cpdf}
\end{eqnarray}
here the quadratic function 
\begin{equation}
Q(\vy_{n,k}) = (\vy_{n,k}-\vmu_{k})^{T}\vSigma^{-1} (\vy_{n,k}-\vmu_k) 
\end{equation}
and $\vSigma$ and $d$ represent the spatial covariance matrix and the number of channels under consideration respectively. In general, multichannel signals analyzed across $K$ segments can have spatio-temporal correlations defined by the  spatio temporal covariance matrix 
\begin{equation}
    \vSigma_{ST}  = \vSigma_{T} \otimes \vSigma 
\end{equation}
here $\vSigma_T$ represents the temporal correlations. In this study, it is assumed that the variations are independent across segments and hence the conditional covariance of $\vy_{n,k}$ reduces to
\begin{eqnarray}
    \vSigma_{ST} = z_k \vSigma 
\end{eqnarray}
Note that in \cite{furui2019scale}, a single channel sEMG signal was modelled and the covariance further reduced to a scalar variance modelled as inverse Gamma random variable. 

In this study, the texture $z_k$ is proposed to follow an exponential distribution    
with the pdf defined as 
\begin{equation}
    p(z_{k})= \frac{1}{\lambda } \exp \bigg({-\frac{z_{k}}{\lambda }}\bigg)
    \label{hiddenpdf}
\end{equation}
where $\lambda$ is a rate parameter. 
The marginal distribution of $\vy_{n,k}$ can be obtained by integrating out the hidden variable $z_{k}$ as follows
\begin{eqnarray}
   p(\vy_{n,k}) &=& \int_{0}^{\infty}p(\vy_{n,k}|z_k) p(z_k) dz_k \nonumber  \\
   & =& \frac{1}{(2\pi)^\frac{d}{2}\lambda \left | \Sigma \right |^\frac{1}{2}}\int_{0}^{\infty}z_{k}^\frac{-d}{2} e^-\Big(\frac{T_{k}^{1}}{z_k}+\frac{z_k}{\lambda }\Big)dz_k
   \label{App_Marginal}
\end{eqnarray}
where $T_k^{1}$ is defined as 
\begin{equation}
    T_k^{1}=\frac{1}{2}\sum_{n=1}^{N}Q(\vy_{n,k})
    \label{T-statistic}
\end{equation}
Using ET II 82(23)a, LET I 146(29) from \cite{gradshteyn2014table} the following integral is identified  
\begin{equation}
\int_{0}^{\infty}x^{\vartheta -1} \exp \bigg(- \frac{A}{x}-Bx\bigg)dx = 2\bigg(\frac{A}{B}\bigg)^{\frac{\vartheta }{2}}K_{\vartheta }\big(2\sqrt{AB}\big)
\label{standard_result}
\end{equation}
where $K(\cdot)$ represents the modified Bessel function of second kind and $\vartheta $ is a order of Bessel function and $A, B$ are its parameters. 
Using (\ref{standard_result}) the marginal distribution (\ref{App_Marginal}) reduces to \cite{eltoft2006multivariate}
\begin{eqnarray}
 p(\vy_{n,k})&=&\frac{2}{(2\pi)^{\frac{d}{2}}\left | \Sigma \right |^{\frac{1}{2}}\lambda }\frac{K_{\frac{d}{2}-1}\Big(\sqrt{\frac{2Q(y_{n,k})}{\lambda }}\Big)}{{\Big(\sqrt{\frac{\lambda Q(y_{n,k}) }{2}}\Big)}^{\frac{d}{2}-1}}\label{Marginal_pdf} 
\end{eqnarray} 
\begin{figure}[H]
        \centering
      \includegraphics[width=0.91\columnwidth]{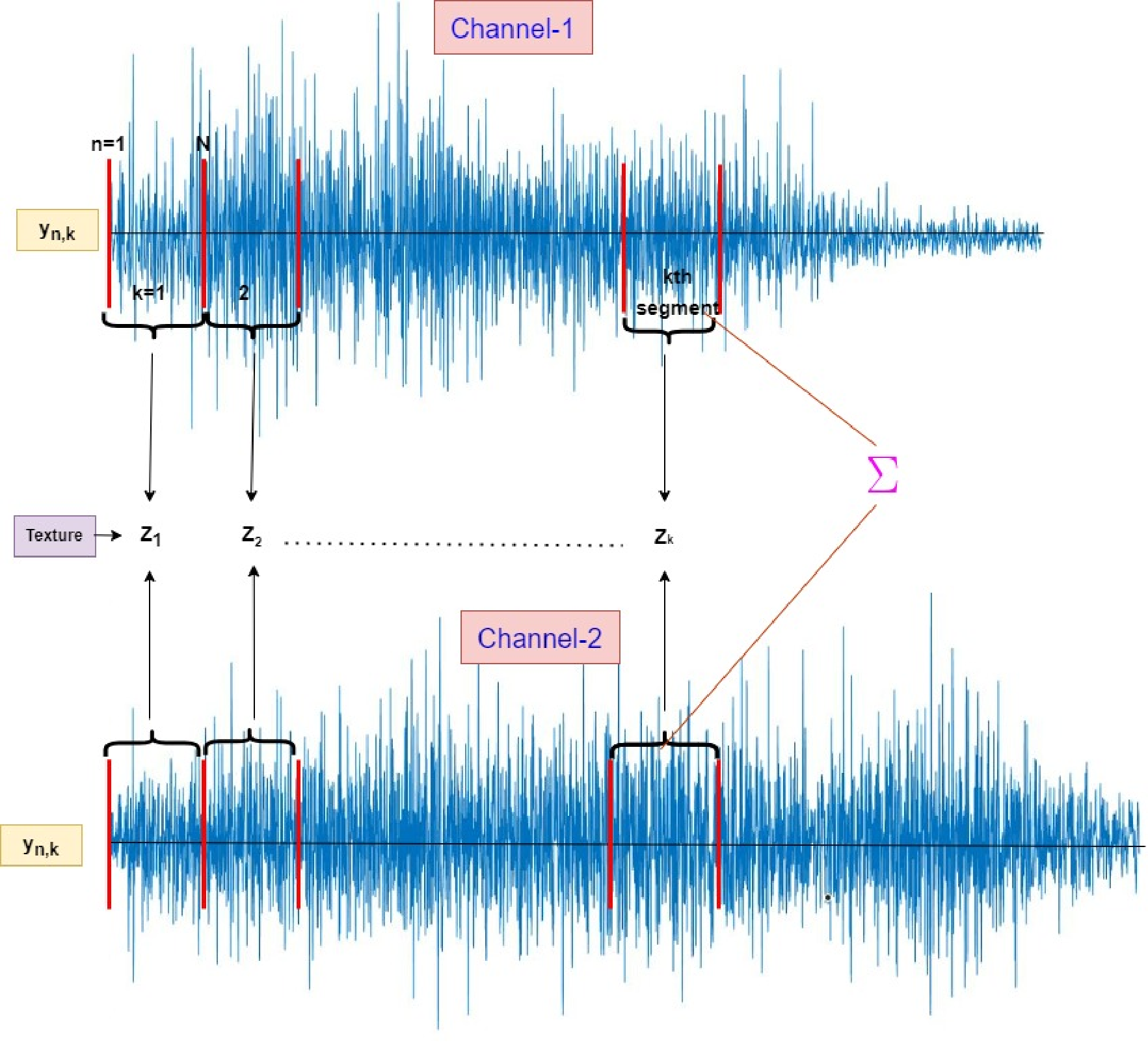}
      \caption{Illustration of two channel sEMG data relation to compound statistical model}
     \label{KN_PLOT}
\end{figure}

\subsection{Estimation Problem}
The complete data likelihood model can be written as 
\begin{equation} \label{pdf:FullLike}
p(\vY,\vz;\vmu,\vSigma,\lambda)= \prod_{k=1}^K \prod_{n=1}^N p(\vy_{n,k}|z_{k};\vmu,\vSigma) p(z_{k};\lambda )
\end{equation} 
where the full observations set $\vY$ is 
\begin{equation}
    \vY = \{\vy_{k}\}_{k=1}^{K}  
\end{equation}
and $\vy_k = \{\vy_{n,k}\}_{n=1}^N$ is the set of $N$ observations within a $k$-th segment,  
the set of texture variables $\vz = \{z_k\}_{k=1}^K$. The parameter set is $\Theta =\{ \lambda ,\vmu, \vSigma \}$ and assumed to be deterministic and unknown. 

The problem of estimation is summarized as follows: given the full data likelihood function (\ref{pdf:FullLike}), a set of measurements $\vY$ that follows the conditional distribution in (\ref{cpdf}), the texture variables $\vz$ assumed to follow the exponential model (\ref{hiddenpdf}), the objective is to estimate the posterior distribution of $\vz$ and the unknown parameters $\Theta$ and assess the estimation performance.
 
\subsection{Parameter estimation using Expectation Maximization (EM) algorithm }
Regarding statistical models involving hidden variables (\ref{pdf:FullLike}), the model parameters are usually estimated using an iterative Expectation Maximization (EM) algorithm \cite{dempster1977maximum}. In this study, {the texture's moments} and the unknown parameters $\Theta$ are estimated using the EM algorithm described in \cite{bishop2006pattern} and variations of the EM for the CG-E are available in \cite{eltoft2006multivariate}. 
It involves the E-step and the M-step as described below. 
\subsubsection{E-step}
In this step, the posterior distribution of the texture $z_k$ is evaluated based on the logarithm of the complete data likelihood model (\ref{pdf:FullLike}).
which is written as 
\begin{eqnarray} \label{Comp_data_LL}
L&=&\sum_{k=1}^{K}\sum_{n=1}^{N}ln\Big(p(y_{n,k}|z_{k})\Big)+\sum_{k=1}^{K}ln\Big(p(z_{k} )\Big)\\ 
&=&-\frac{NKd}{2}ln(2\pi)-\frac{Nd}{2}\sum_{k=1}^{K}ln(z_{k})-\frac{NK}{2}ln(\left | \vSigma \right |) 
\nonumber \\
 &&
-\frac{1}{2}\sum_{k=1}^{K} \sum_{n=1}^{N}\frac{Q(\vy_{n,k})}{z_{k}}-Kln(\lambda )-\frac{1}{\lambda }\sum_{k=1}^{K}z_{k}
\nonumber 
\label{LCDL}
\end{eqnarray}
Assuming the texture variables $z_k$ are independent across segments, at each iteration, the posterior pdf of $\vz$ can be approximated as the product of individual posteriors as 
\begin{equation}
q_{j}(\textbf{z})=\prod_{k=1}^{K}q_{j}(z_{k})
\end{equation}
Gathering the $z_k$ terms in (\ref{Comp_data_LL}), the log posterior of $z_k$ is obtained as
\begin{equation}
    ln(q(z_k))=-\frac{Nd}{2}ln(z_k)-\frac{T_k^{1}}{z_k}-\frac{z_k}{\lambda }+ C
\end{equation}
}
Applying the exponential function on both the sides leads to
\begin{equation}
    q(z_k)\propto z_k^{-\frac{Nd}{2}}e^{-\frac{T_k^{1}}{z_k }-\frac{z_k}{\lambda }}
\end{equation}
The normalization constant for above equation is evaluated as:
\begin{equation}
    V=\int_{0}^{\infty}q(z_k)dz_k
\end{equation}
using (\ref{standard_result}) $V$ becomes
\begin{eqnarray}
V&=&\int_{0}^{\infty} z_k^{-\frac{Nd}{2}}e^{-\frac{T_k^{1}}{z_k }-\frac{z_k}{\lambda }}dz_k,\\
\nonumber
&=&2\Big(\frac{T_k^{1} }{1/\lambda}\Big)^{\frac{\varsigma}{2} }K_\varsigma \Big(2\sqrt{T_k^{1}/\lambda}\Big)\\
\end{eqnarray}
where $\varsigma$ is a order of the Bessel function and is defined as
\begin{equation}
    \varsigma=\frac{-Nd}{2}+1
\end{equation}
The posterior distribution of $z_k$ is given by the pdf 
\begin{equation}
    q(z_k)=\frac{z_k^{-\frac{Nd}{2}}e^{-\frac{T_k^{1}}{z_k }-\frac{z_k}{\lambda }}}{2\Big(\frac{T_k^{1} }{1/\lambda}\Big)^{\frac{\varsigma}{2} }K_\varsigma \Big(2\sqrt{T_k^{1}/\lambda}\Big)}
    \label{posterior distribution}
\end{equation}
The estimated posterior (\ref{posterior distribution}) is not a known distribution. 
However, expectations of functions $h(z_k)$ required in the following M-step can be evaluated as follows 
\begin{eqnarray}
 \left\langle  h(z_{k})\right\rangle &=& \int_{0}^{\infty} h(z_{k}) q(z_{k}) dz_{k}
\end{eqnarray}
By substituting for $q(z_k)$ from (\ref{posterior distribution}) and utilizing (\ref{standard_result}), the posterior mean \cite{8450847} is obtained as 
\begin{eqnarray}
\left\langle z_{k} \right\rangle&=&\sqrt{\Big(\frac{T_k^{1} }{1/\lambda}\Big)}\frac{K_{\varsigma +1}(2\sqrt{T_k^{1}/\lambda})}{K_{\varsigma}(2\sqrt{T_k^{1}/\lambda})}
\label{Eofz}
\end{eqnarray}
Similarly the other required moments $\left\langle  \ln (z_k)\right\rangle$, $\left\langle  \frac{1}{z_{k}}\right\rangle$ are evaluated as \\
\begin{eqnarray}
\left\langle \ln z_{k} \right\rangle&=&\frac{1}{2}ln\Big(\frac{T_k^{1}}{1/\lambda  }\Big)+\frac{\left.\frac{\partial K_{\xi}(2\sqrt{T_k^{1}/\lambda})}{\partial \xi}\right|_{\xi=\varsigma}}{K_{\varsigma}(2\sqrt{T_k^{1}/\lambda})}\\
\left\langle \frac{1}{z_{k}} \right\rangle&=&{\Big(\frac{T_k^{1} }{1/\lambda}\Big)}^{-\frac{1}{2}}\frac{K_{\varsigma -1}(2\sqrt{T_k^{1}/\lambda})}{K_{\varsigma}(2\sqrt{T_k^{1}/\lambda})}
\label{Elogzinvz}
\end{eqnarray}.
\subsubsection{M-step}
In this step, the expectation of log likelihood function $ L$ is.


\begin{multline}
     \left\langle  \ln  L \right \rangle_{q(z_k)}=-\frac{Nd}{2}\sum_{k=1}^{K}\left\langle  ln z_{k}\right\rangle-\sum_{k=1}^{K}\left\langle \frac{1}{z_{k}} \right\rangle T_k^{1}- \frac{NK}{2}ln(\left | \vSigma \right |) \\
    -Kln(\lambda )-\frac{1}{\lambda }\sum_{k=1}^{K}\left\langle  z_{k}\right\rangle
    \label{M-step:exp}
\end{multline}
The parameters are estimated by maximizing the expectation as 
\begin{equation}
    \nabla_{\mu,\Sigma,\lambda }  \left\langle  \ln  (L) \right \rangle_{q(z_k)}  =0
    \label{M_step:maximize}
\end{equation}

Using the moments from (\ref{Eofz}) to (\ref{Elogzinvz}) in (\ref{M_step:maximize}), the parameters are estimated as 
\begin{eqnarray}
\hat{\lambda }&=&\frac{1}{K}\sum_{k=1}^{K}\left\langle  z_{k}\right\rangle,\label{updated_parameters1}\\
\hat{\vmu}_k&=&\frac{\sum_{k=1}^{K}\bar{y}_{k}\eta _{k}}{\sum_{k=1}^{K}\eta _{k}}, \label{updated_parameters2}\\
\hat{\vSigma}&=&\frac{1}{NK}\sum_{k=1}^{K}\sum_{n=1}^{N}\eta _{k} Q(\vy_{n,k})
\label{updated_parameters3}
\end{eqnarray}
Where  $\bar{y}_{k}=\frac{1}{N}\sum_{n=1}^{N}y_{n,k}$ and $\eta_{k}=\left\langle  \frac{1}{z_k}\right\rangle$.
\subsubsection{Convergence criteria}
The E and M steps are repeated until the convergence criterion defined below is satisfied. $\phi^{i}$ represents the  sum of absolute change in the consecutive parameter estimates at the $i^{th}$ iteration and defined as
\begin{equation}
    \phi^{(i)}=\left | \lambda ^{i} - \lambda ^{i-1}\right |+\left | \mu_{k}  ^{i} - \mu_{k} ^{i-1}\right |+\left | \Sigma_{c} ^{i} - \Sigma_{c} ^{i-1}\right |
    \label{convergence1}
\end{equation}
When the change $\phi^{(i)}$ becomes sufficiently small ie,.
\begin{equation}
    \phi^{(i)} \leq \phi_o
    \label{convergence2}
\end{equation}
the iterations are halted. Here $\phi_o$ represents a pre-defined value $10^{-5}$. 
The EM-algorithm is summarized in Alg. 1 .

\begin{algorithm}[h] \label{alg:cap}
\hrulefill

Algorithm 1: EM-algorithm for CG-E model

\hrulefill

~\textbf{Input:} ~~Measurements $Y$ \\
~1. Initialize the parameters $ \lambda_{0},\mu_{0}, \Sigma_{0}$ in (\ref{posterior distribution}) \\
~2. Set $i = 1$ \\
~3. \textbf{while} $i \leq 
  I_{max}$ \textbf{do}  \\
~~~~~(a) \textbf{E-step:}\\
~4. ~~~~~~~\textbf{for} {$j = 1 \rightarrow J$} \textbf{do}{ \\
~~~~~~~~~~~~Update: \\
~5. ~~~~~~~~the statistic $T_k^{1}$ in (\ref{T-statistic})\\
~6. ~~~~~~~~moments $\left\langle z_{k} \right\rangle$, $\left\langle \frac{1}{z_{k}}\right\rangle$ and $\left\langle lnz_{k} \right\rangle$ in (\ref{Eofz}) to (\ref{Elogzinvz})  
   } \\
~7. ~~~~~~~\textbf{end for}   \\
~~~~~(b) {\textbf{M-Step:} \\
~8.
~~~~~~~~~Update parameters $\lambda, \mu, \vSigma$ in (\ref{updated_parameters1}) to (\ref{updated_parameters3})\\
~~~~~(c) \textbf{Stopping Criterion:} \\ {
~9.~~~~~~~Update  $\phi^{i}$ from (\ref{convergence1})  \\
10.~~~~~~~~~~~\textbf{if}  condition in (\ref{convergence2}) is true { \\
11.~~~~~~~~~~~Convergence reached, stop.} \\
12.~~~~~~~~~~~\textbf{else} { \\
13.~~~~~~~~~~~Set $ i \leftarrow i + 1$ \\
14.~~~~~~~~~~~Repeat steps (a), (b) and (c) \\ }
15.~~~~~~~~~~~\textbf{end if} \\
 } }
16. \textbf{end while}  \\
\textbf{Output:} Estimated posterior distribution $z_{k}$ and ML estimates of $\lambda, \mu, \Sigma$ \\
\hrulefill
\label{Algo}
\end{algorithm}

\subsection{Comparison Models}
In this work, we compare our proposed model (CG-E) with the benchmark \cite{furui2021emg}, compound Gaussian distribution with inverse gamma texture (CG-IG) and another model, compound Gaussian distribution with the Gamma texture (CG-G) {\cite{wang2006maximum}. 
\subsubsection{CG-IG Model}
The texture $z_k$ follows an inverse gamma distribution 
\begin{equation} \label{InvGamTexture}
    p(z_k) = \mathcal{IG}\big(z_k; \alpha_{IG}, \beta_{IG}\big)
\end{equation}
and the conditional distribution of $y_{n,k}$ given texture $z_{k}$ is the same  as (\ref{cpdf}) with the parameters $\vmu_{IG},\vSigma_{IG}$. Note that this benchmark model is based  on the scale mixture model in \cite{furui2019scale}. Let $\Theta_1=\left \{ \vmu_{IG},\vSigma_{IG},\alpha_{IG} ,\beta_{IG}  \right \}$ be the parameter set for the CG-IG model. The EM-algorithm for estimation of the parameters is summarized as follows, similar results can be found in \cite{furui2019scale, furui2021emg}.
\begin{itemize}
    \item E-step:
    The posterior distribution of $z_k$ has a closed form expression, which follows inverse gamma distribution ie,.
\begin{equation}
    q(z_k) = \mathcal{IG}\big(z_k; \alpha^*_{IG}, \beta^*_{IG}\big)
    \label{posterier_IG}
\end{equation}
\end{itemize}
where $\alpha^{*}_{IG}$ and $\beta^{*}_{IG}$ are written as
\begin{eqnarray}
     \alpha^{*}_{IG} &=&\frac{Nd}{2}+\alpha_{IG} \\
     \beta^{*}_{IG} &=&\beta_{IG}+T^1_{k} \nonumber
\end{eqnarray}
The moments of $z_k$ are given as follows:
\begin{eqnarray}  \label{moments_z_cgig}
     \left\langle \ln z_{k} \right\rangle&=&\ln(\beta^{*}_{IG} )-\psi (\alpha^{*}_{IG})\\
     \left\langle \frac{1}{z_k} \right\rangle&=&\frac{\alpha^{*}_{IG} }{\beta^{*}_{IG} } \nonumber 
\end{eqnarray}
\begin{itemize}
    \item M-step:
    The estimates of ${\Theta_1 }$ are obtained as follows:
  An estimate of  $\alpha_{IG}$ is found by solving the 
 non-linear equation 
 \begin{equation}
     K\psi (\alpha_{IG} )-K \ln(\beta_{IG} )+\sum_{k=1}^{K}\left\langle \ln{(z_k)} \right\rangle=0
     \label{non-linear}
 \end{equation}
 using the Newton Raphson Method \cite{kelley2003solving} and by using the solution of (\ref{non-linear}) an estimate of  ${\beta_{IG} }$ is obtained as
 \begin{equation}
\breve{\beta}_{IG}=\frac{K\breve{\alpha}_{_{IG}}}{\sum_{k=1}^{K}\left\langle \frac{1}{z_{k}} \right\rangle}
     \label{beta_est}
 \end{equation}
 The remaining estimates $\vmu_{IG},\vSigma_{IG}$ are similar to (\ref{updated_parameters2}) and (\ref{updated_parameters3}) except for the moments of $z_k$ are replaced by those in (\ref{moments_z_cgig}). 
\end{itemize}

 \subsubsection{CG-G Model}
 Here the texture $z_k$ is considered as a gamma random variable.
 \begin{equation} \label{GammaTexture}
     p(z_k) = \mathcal{G}(z_k; \alpha_G, \beta_G) 
 \end{equation}
 The conditional distribution of $y_{n,k}|z_{k}$ is again similar to (\ref{cpdf}) with the parameters $\mu_G, \vSigma_G$. $\alpha_G, \beta_G$ are the parameters
 of the gamma distribution. Let $\Theta_2=\left \{ \vmu_{G},\vSigma_{G},\alpha_{G} ,\beta_{G}  \right \}$ be the parameter set of CG-G model. Note that this model is used for modeling non-stationary radar clutter \cite{gini2000performance}. 
 \begin{itemize}
     \item E-step:
     The posterior distribution of $z_k$ 
with CG-G model is similar to (\ref{posterior distribution}) and is given as 
\begin{equation}
    q(z_k)=\frac{z_k^{\alpha_{G} -\frac{Nd}{2}-1}e^{-\frac{T_k^{1}}{z_k }-\beta_{G} z_{k}}}{2\Big(\frac{T_k^{1} }{\beta_{G}}\Big)^{\frac{\nu}{2} }K_{\nu} \Big(2\sqrt{\beta_{G} T_k^{1}}\Big)}
\end{equation}
here $\nu = -\frac{Nd}{2}+\alpha_G$ represents the order of the Bessel function. The moments of $z_{k}$ are similar to those of CG-E model (\ref{Eofz})  to (\ref{Elogzinvz}) with the following replacements. 
\begin{eqnarray} \label{CGG-replacements}
    \frac{1}{\lambda} &\rightarrow & \beta_G \nonumber \\
    \varsigma &\rightarrow & \nu 
\end{eqnarray}

  \item M-step: 
  The estimate of ${\alpha}_{G} $ is obtained by solving the non-linear equation given below
     \begin{equation}
          K\psi (\alpha_{G} )-Klog(\beta_{G} )-\sum_{k=1}^{K}\left\langle \ln{(z_k)} \right\rangle=0
          \label{non-linear1}
     \end{equation}
 \end{itemize}
 using (\ref{non-linear1}), the estimate of ${\beta_{G} }$ is obtained as
 \begin{equation}
     \breve{\beta}_{G}=\frac{K\breve{\alpha}_{_{G}}}{\sum_{k=1}^{K}\left\langle {z_{k}} \right\rangle}
 \end{equation}
 The estimates of $\vmu_{G},\vSigma_{G}$ are similar to (\ref{updated_parameters2}) and (\ref{updated_parameters3}) except for the modified moments of $z_k$.}

\subsection{Evaluation methods}
\textbf{{Visual inspection}\cite{spanos2019probability}}:
It is a graphical approach to visualise the level of agreement between the histogram based empirical pdf (empdf) and an estimated pdf. In this study, these estimated compound pdfs  are based on the CG-E, CG-G and CG-IG models.

\textbf{Moment Analysis}: In this analysis, the statistical moments estimated from the three models are compared with those of the empdf. 
The requried moments are computed from the following 
\begin{equation} \label{MultiMoments}
    E(h(\vY))  = \int h(\vY) p(\vY, \vz) d \vY d\vz  
\end{equation}
For the three models, the closed form joint pdfs lead to closed form moments. The moments from these models are compared with the data based moments which are evaluated numerically from (\ref{MultiMoments}) by replacing the joint pdf with the numerical empdf. 

\textbf{Kullback–Leibler divergence (KLD) \cite{kullback1997information}}: 
 It is a statistical metric used to measure the distance between two pdfs. Let $q_{1}$ and $q_{2}$ be the empdf and the estimated model respectively, then the KLD between them is evaluated as 
\begin{equation}
D_{KL}(q_1 || q_2) =\sum_{x}\sum_{y}q_1(x,y) \ln \bigg(\frac{q_1(x,y)}{q_2(x,y)} \bigg)
\end{equation}\\
If {these} models match with each other then the $D_{KL}(q_1 || q_2)$ equals $0$. Thus a lower $D_{KL}(p_1 || p_2)$ indicates that an  estimated model is closer to the empdf.

\textbf{Coefficient of determination (COD) R-squared\cite{cohen1983applied}}: 

It is a statistical measure that determines how well the estimated model fits the empdf. Specifically, it quantifies how much of the overall variance, the estimated model can explain. As the value of the $R^{2}$ approaches $1$, the agreement between the estimated model and the empdf improves.


\textbf{Log-Likelihood Values (LLV)\cite{spanos2019probability,pawitan2001all}} : 

The LLV is another measure to compare two different statistical models. In order to determine which of the models is statistically significant, the likelihood values associated with the models are evaluated separately and compared. We determine the LLV for the three models mentioned above.

\begin{figure*}[t]
 \captionsetup[subfigure]{justification=centering}
    \centering
      \begin{subfigure}{0.31\textwidth}
        \includegraphics[width=\textwidth]{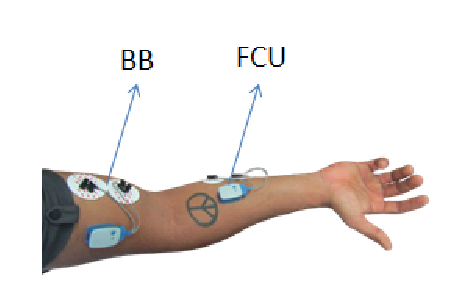}
          \caption{}
          \label{electrode placement}
      \end{subfigure}
      \begin{subfigure}{0.20\textwidth}
        \includegraphics[width=\textwidth]{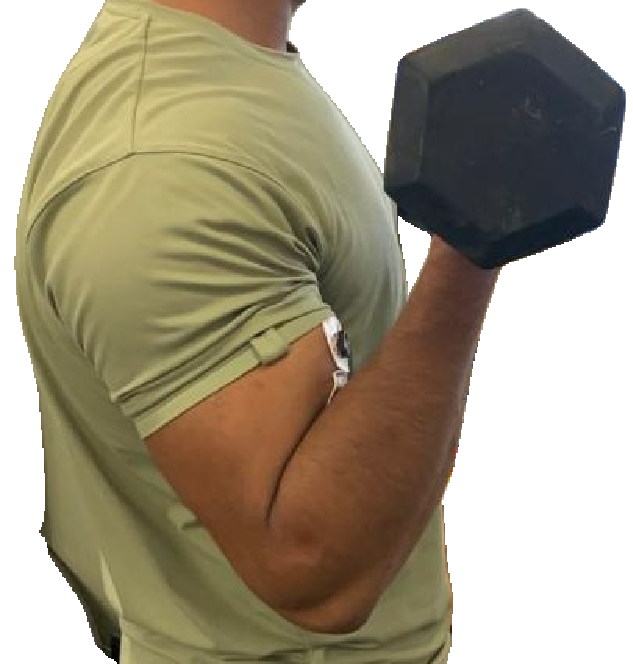}
          \caption{}
          \label{TONIC}
      \end{subfigure}
      \begin{subfigure}{0.26\textwidth}
        \includegraphics[width=\textwidth]{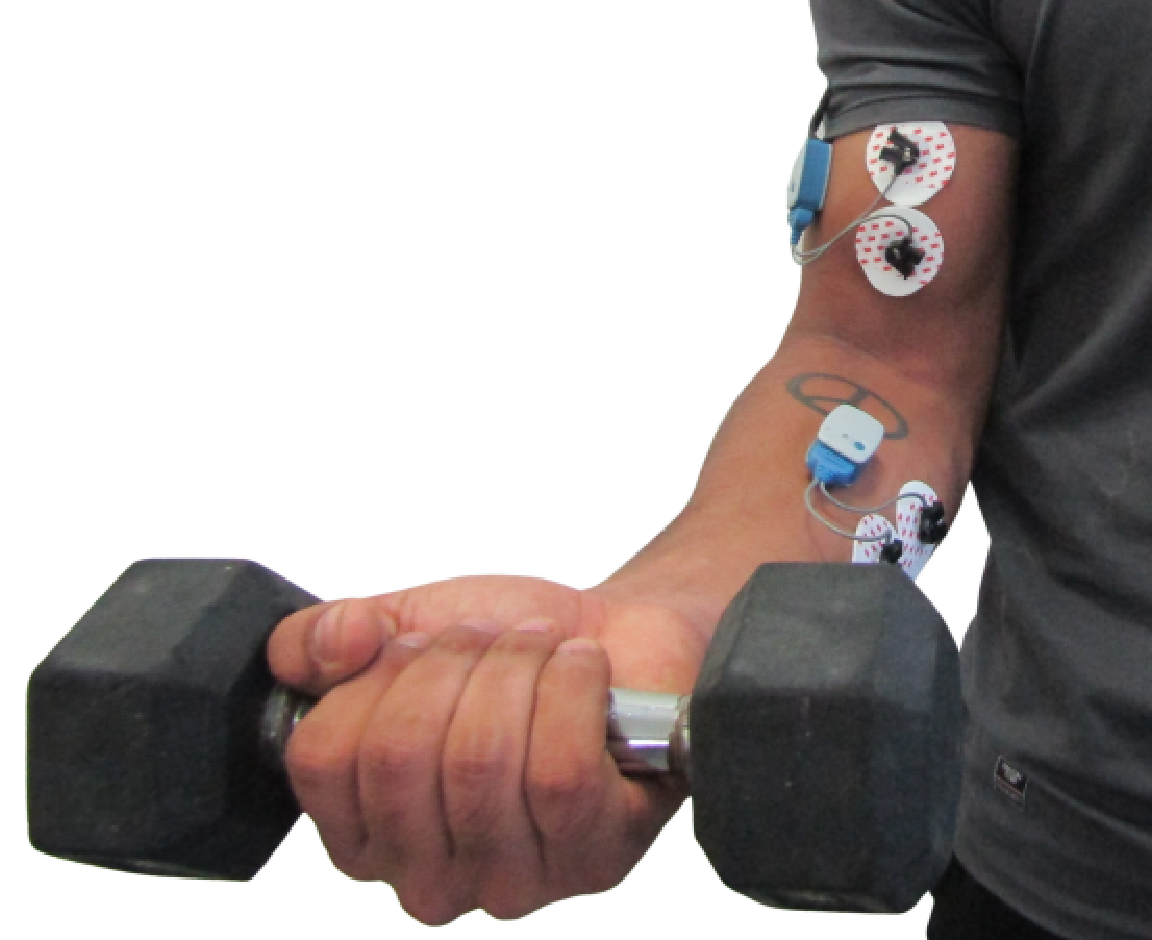}
          \caption{}
          \label{METRIC}
      \end{subfigure}
    
      \caption {(a) Placement of electrodes  on the BB and FCU during weight training. (b) Isotonic activity: Performing bicep curls (c) Isometric activity: Holding the dumbbell at 90 $^{\circ}$.  }
      \label{setup}
\end{figure*}

\begin{table*}[t]
\centering
\caption{Anthropometrics of Participants } 
\begin{tabular}{cccccc}
\hline 
\textbf{Subject} & \textbf{COBB*(inches)} & \textbf{COFCU*(inches)} & \textbf{Experience}             & \textbf{Weight (kg)}              & \textbf{Height (cms)}                  \\ \hline
1      &  10.5                  &10                  & No                             & 58 &  175                                 \\ \hline
2       & 11.5                  &  10.5               & No                              &  75 &  183                                  \\ \hline
3       & 13                  &  10.5                & No                              &  70 & 173.7  \\ \hline
4               &  12.8                  &  10.5                &  2 months  &  81  &  182    \\ \hline
5               &  12.5                  &  10               &  3 months  &  63  &  175    \\ \hline
6                &  11                    &  9.8                 &  3 months   & 57                         & 174                            \\ \hline
7                &  12                    &  10                  &  4 months & 75                         & 182                           \\ \hline
8                & 13.8                                         & 11.9                                       &  1 year &  65  &  173     \\ \hline
9                &  14.3                  &  12                  &  2 years  &  77  &  176     \\ \hline
10                &  13.8                  &  12.2                &  1 year   &  79  &  182.8  \\ \hline 

\end{tabular}
\\ \vspace{0.2cm}
{* COBB and COFCU stand for circumference of BB and FCU respectively}
\label{details_participants}
\end{table*}

\begin{table}[t]
\centering
\caption{Characteristics of EMAHA-DB2 dataset}
\begin{tabular}{cccccccc}
 \hline
Weight             &  0kg to 10kg & \\ \hline
Muscles        & \multicolumn{5}{c}{BB and FCU}  \\ \hline
Subjects       & \multicolumn{5}{c}{10}                  \\ \hline
Rest duration          & \multicolumn{5}{c}{10sec}                              \\ \hline
Activity duration      & \multicolumn{5}{c}{8sec}                              \\ \hline
No of repetitions        & \multicolumn{5}{c}{09}                               \\  \hline
sEMG sensor            & \multicolumn{5}{c}{Noraxon}                          \\  \hline
Electrode              & \multicolumn{5}{c}{Agcl} \\ \hline
Sampling frequency(Hz) & \multicolumn{5}{c}{2000}                             \\ \hline
No of channels         & \multicolumn{5}{c}{2}  \\
\hline 
\end{tabular} 
\label{tab:EMAHA-DB2}
\end{table}

\section{Data Description}

In this work, a novel sEMG dataset termed electromyographic analysis of human arm activities - database 2 (EMAHA-DB2) is developed. Ten healthy participants aged between $18-21$ years were selected based on three levels of strength training experience: a) Beginner - with no prior training experience, b) Intermediate - with a few weeks of training experience and c) Trained - with at least one year of training experience \cite{santos2021classification}. Participants were free from all muscle disorders for the past one month prior to data collection. Prior to participating in the experiment, the purpose of the study was explained and an informed consent was obtained from the subjects. The data collection procedure was approved by the institutional ethics committee of the Indian Institute of Information Technology Sri City (No. IIITS/EC/2022/01) dated 19 September 2022 as per the principles of the Declaration of Helsinki.  
Before data acquisition session, the surface of the skin at the muscle site under consideration is cleaned with an alcohol based wipe to reduce the skin impedance. In EMAHA-DB2, sEMG signals are acquired using the Noraxon's Ultium sensors. As shown in fig. \ref{setup}(a), Ultium sensors are placed at two muscle sites 1) biceps brachii (BB) representing the upper arm activity and 2) flexor carpi ulnaris (FCU) representing the forearm activity. Signal acquisition characteristics of the sensor are: $16$- bit A/D; Sampling rate: $2000$ samples/sec; cutoff frequency: $20-450$ Hz. The weights used during the activity include $0$kg, $1$kg, $2.5$kg, $5$kg, $6$kg, $9$kg and $10$kg. During the measurement, the subject is in a standing position and the weight is placed on a table at a convenient height. Each activity has three phases  1) rest ($10$s), 2) action ($5$s) and 3) release ($3$s) with a total duration of $18$s. Each activity is repeated nine times. In order to avoid muscle fatigue, subjects rest for two minutes between different activities. Further details of experiments are given below.  The anthropometric details of the participants are shown in the table  \ref{details_participants} and a summary of the dataset is presented in table \ref{tab:EMAHA-DB2}. 
The EMAHA-DB2 dataset is available \href{http://doi.org/10.7910/DVN/MBG5UY}{here}.

\subsubsection{Experiment-I}
 {
In the first experiment as shown in fig. \ref{setup}(b), the subjects were asked to perform bicep curls with the right arm using the seven weights mentioned above. Recall that the biceps curl corresponds to isotonic muscle contractions \cite{mayhew1995muscular}. } 
\subsubsection{Experiment-II}
{
In this experiment, as shown in fig \ref{setup}(c), the subjects were asked to hold a dumbbell with the right hand at 90$^{\circ}$ with respect to the upper arm i.e., the dumbbell is held in the transverse plane with its axis parallel to the frontal axis. The same set of weight variations from experiment $I$ are used.     
{Recall, for holding a weight, the arm flexion corresponds to isometric contractions \cite{baley1966effects}.} }

\begin{figure*}[btp]
\captionsetup[subfigure]{justification=centering}
    \centering
      \begin{subfigure}{0.31\textwidth}
        \includegraphics[width=\textwidth]{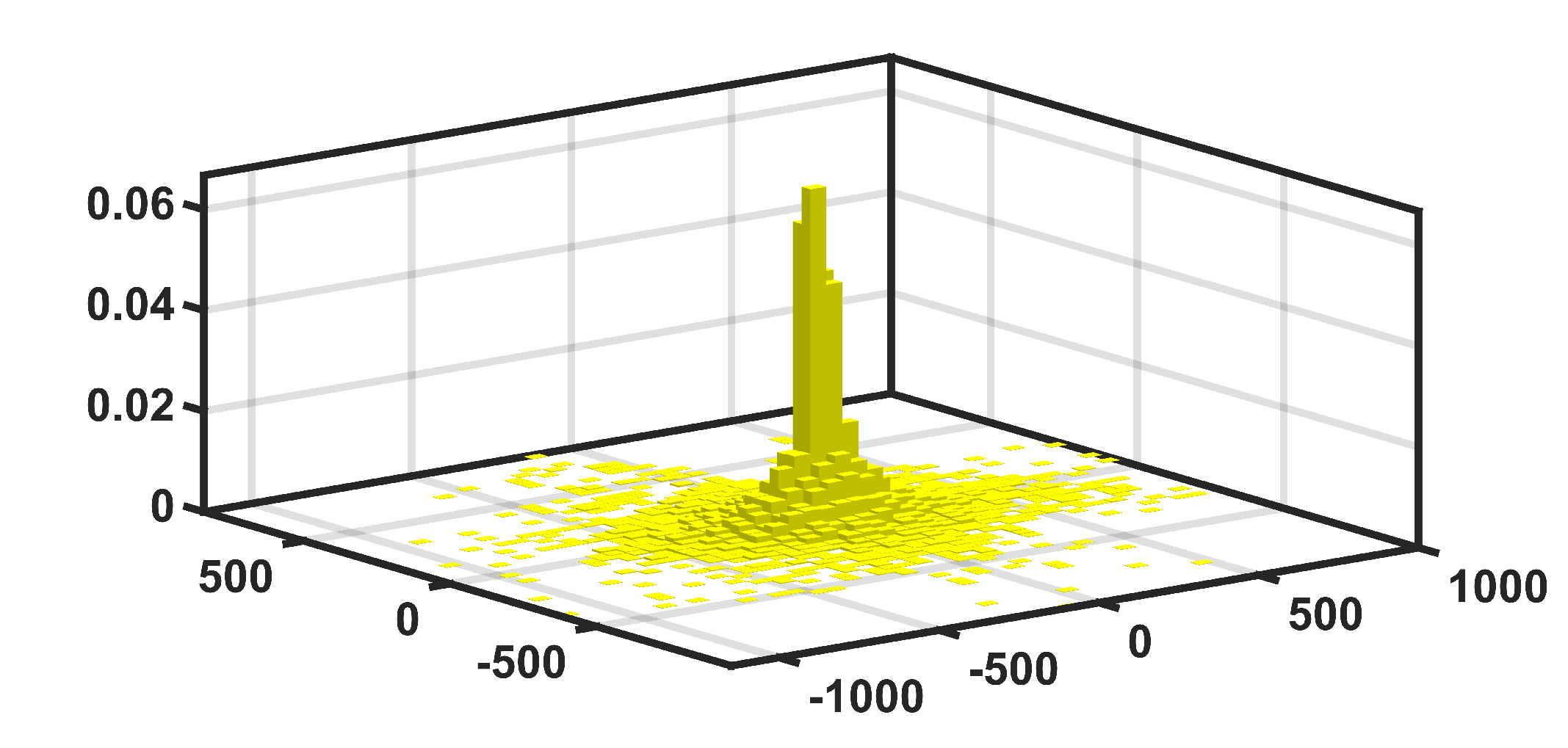} 
          \caption{}
          \label{mpdf_isot}
      \end{subfigure}
      \begin{subfigure}{0.31\textwidth}
        \includegraphics[width=\textwidth]{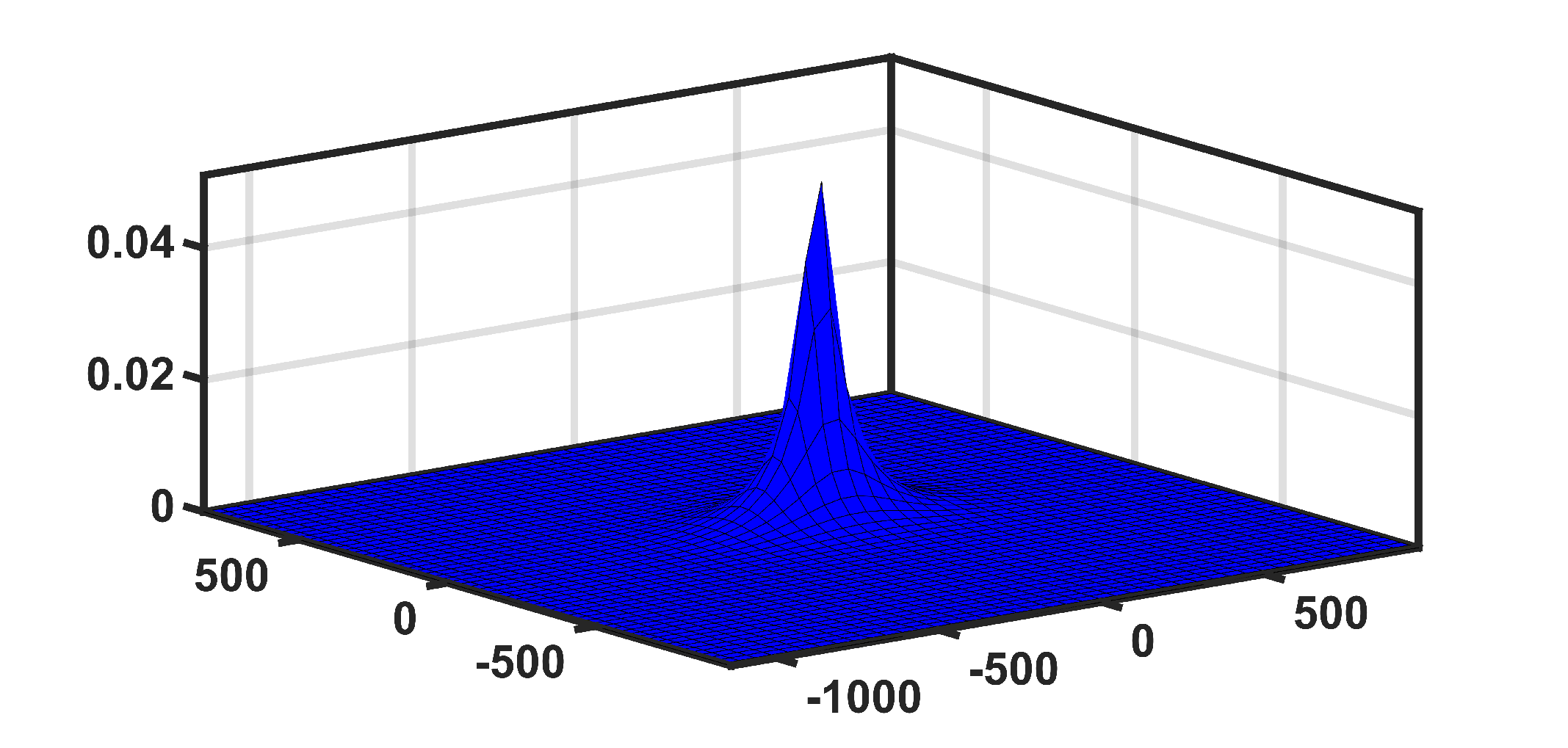} 
          \caption{}
          \label{mpdf_isot_Exp}
      \end{subfigure}
     
      \begin{subfigure}{0.31\textwidth}
        \includegraphics[width=\textwidth]{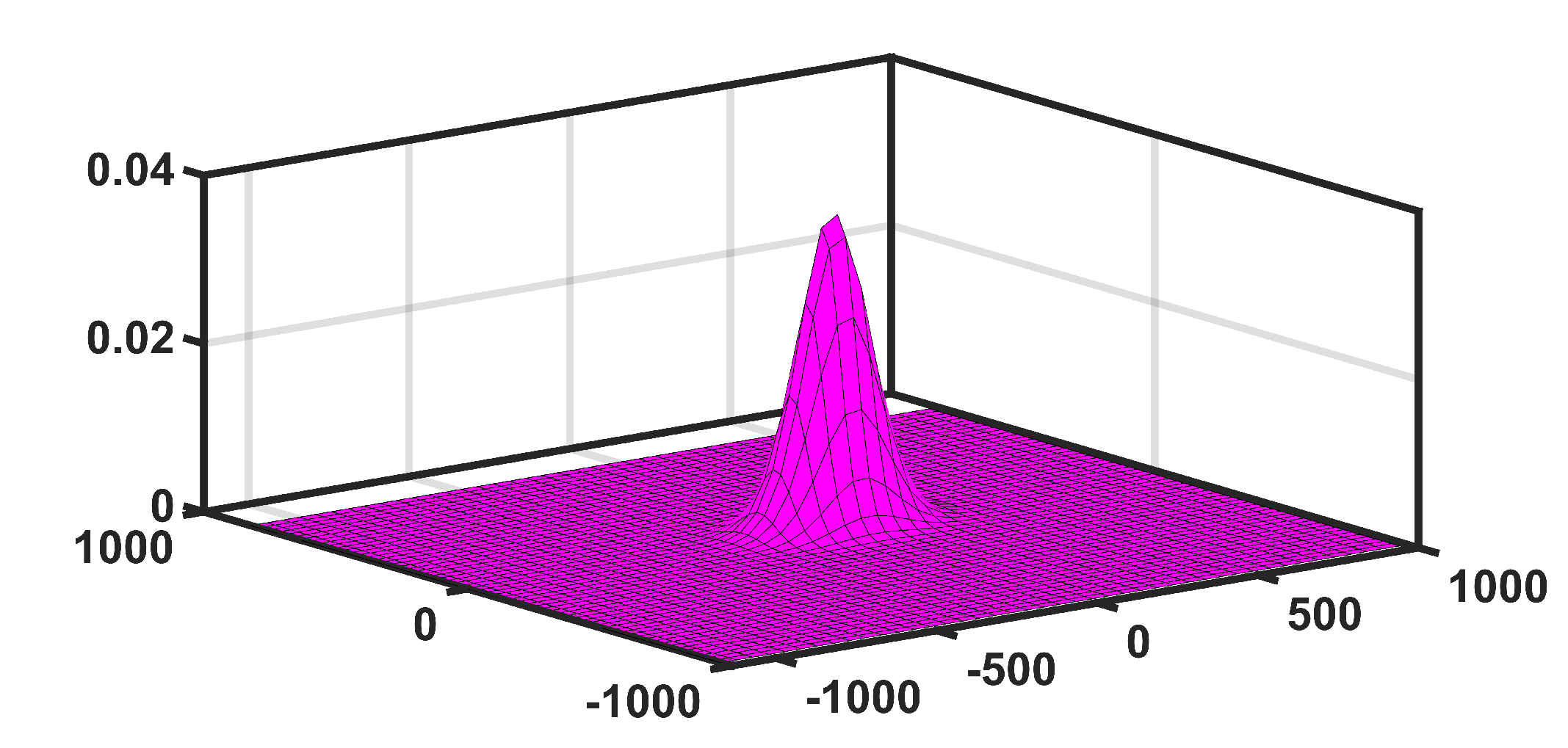} 
          \caption{}
          \label{mpdf_isot_Gamma}
      \end{subfigure}
      \begin{subfigure}{0.31\textwidth}
        \includegraphics[width=\textwidth]{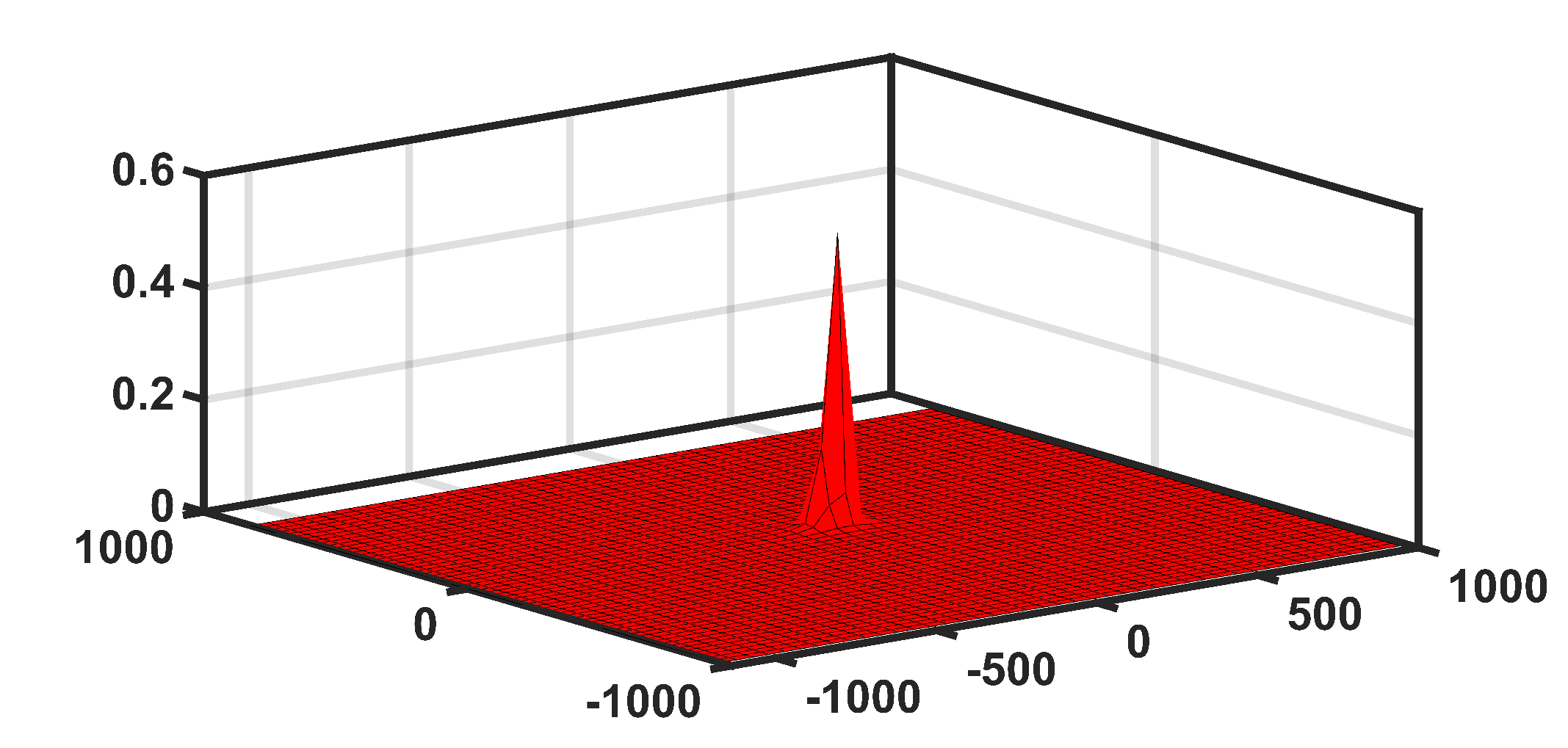}
          \caption{}
          \label{mpdf_isot_SMM}
      \end{subfigure}
      
     \caption {Visual comparisons between (a) empdf (yellow) and estimated pdfs from models: (b) CG-E (blue), (c) CG-G (magenta) and (d) CG-IG (red) for isotonic activity during $6$ kg lifting corresponding to the subject-1 and trial-8. }
      \label{fig:VisualComp}
\end{figure*}

\begin{figure*}[t]
\captionsetup[subfigure]{justification=centering}
    \centering
      \begin{subfigure}{0.31\textwidth}
        \includegraphics[width=\textwidth]{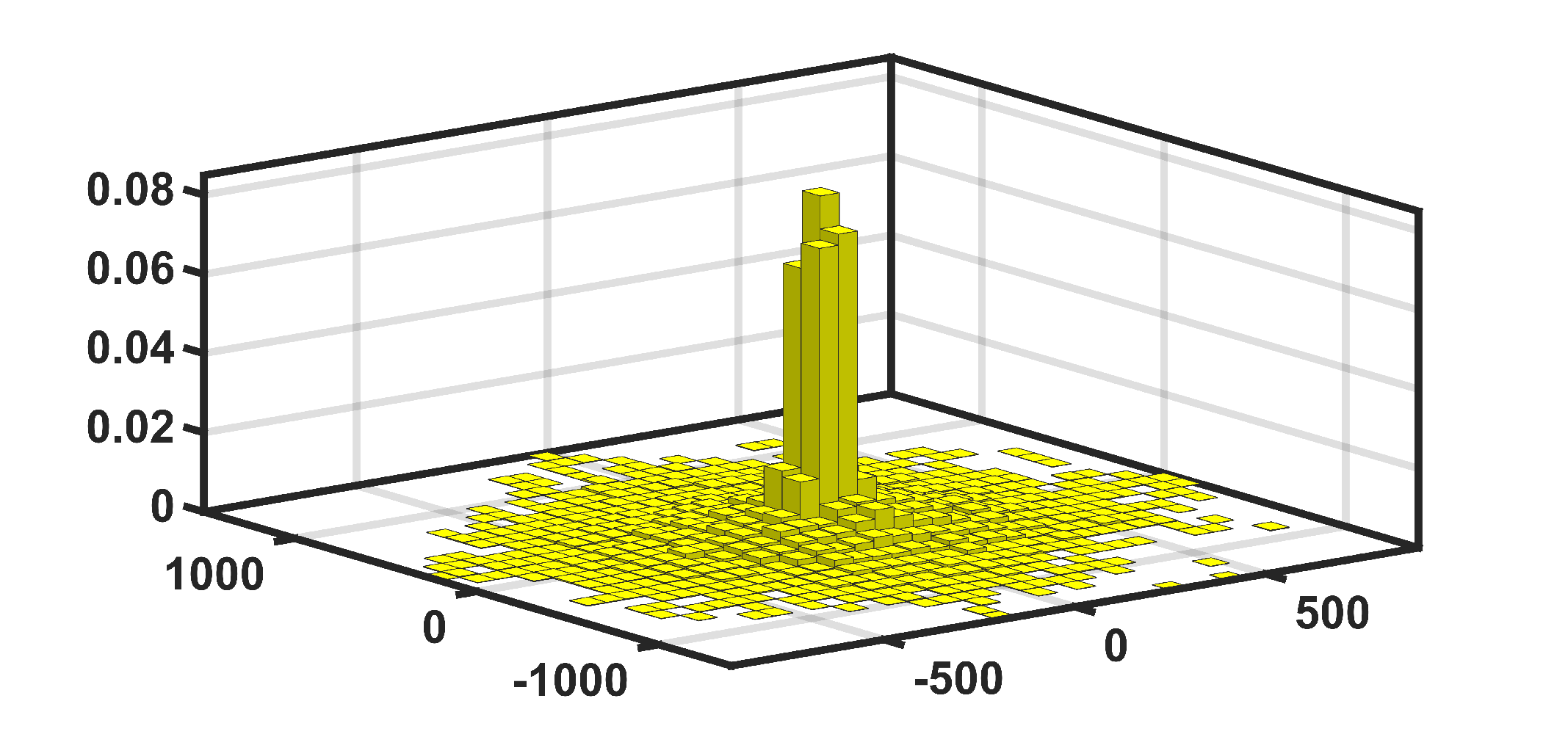}
          \caption{}
          \label{mpdf_isom}
      \end{subfigure}
      \begin{subfigure}{0.31\textwidth}
        \includegraphics[width=\textwidth]{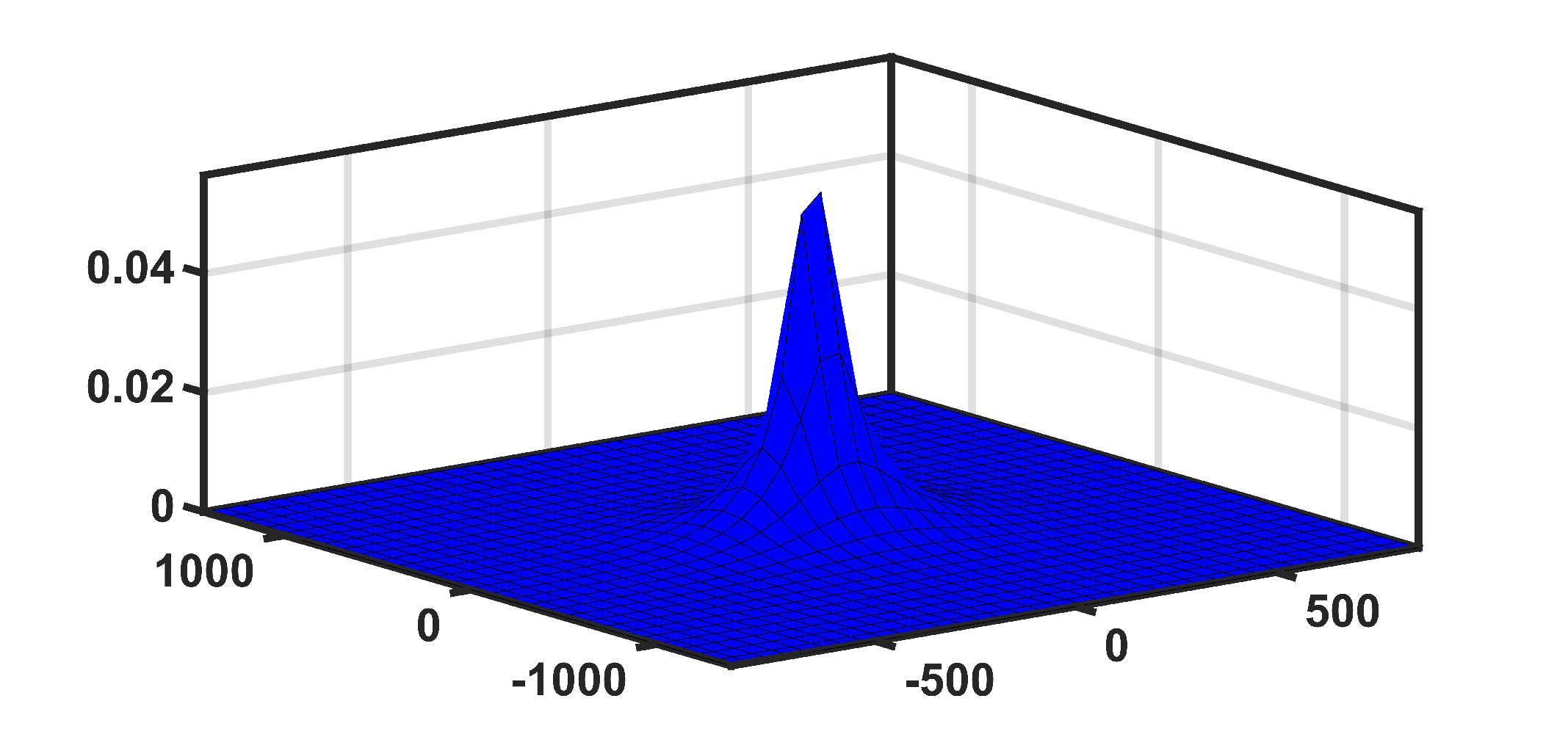}
          \caption{}
          \label{mpdf_isom_Exp}
      \end{subfigure}
     
      \begin{subfigure}{0.31\textwidth}
        \includegraphics[width=\textwidth]{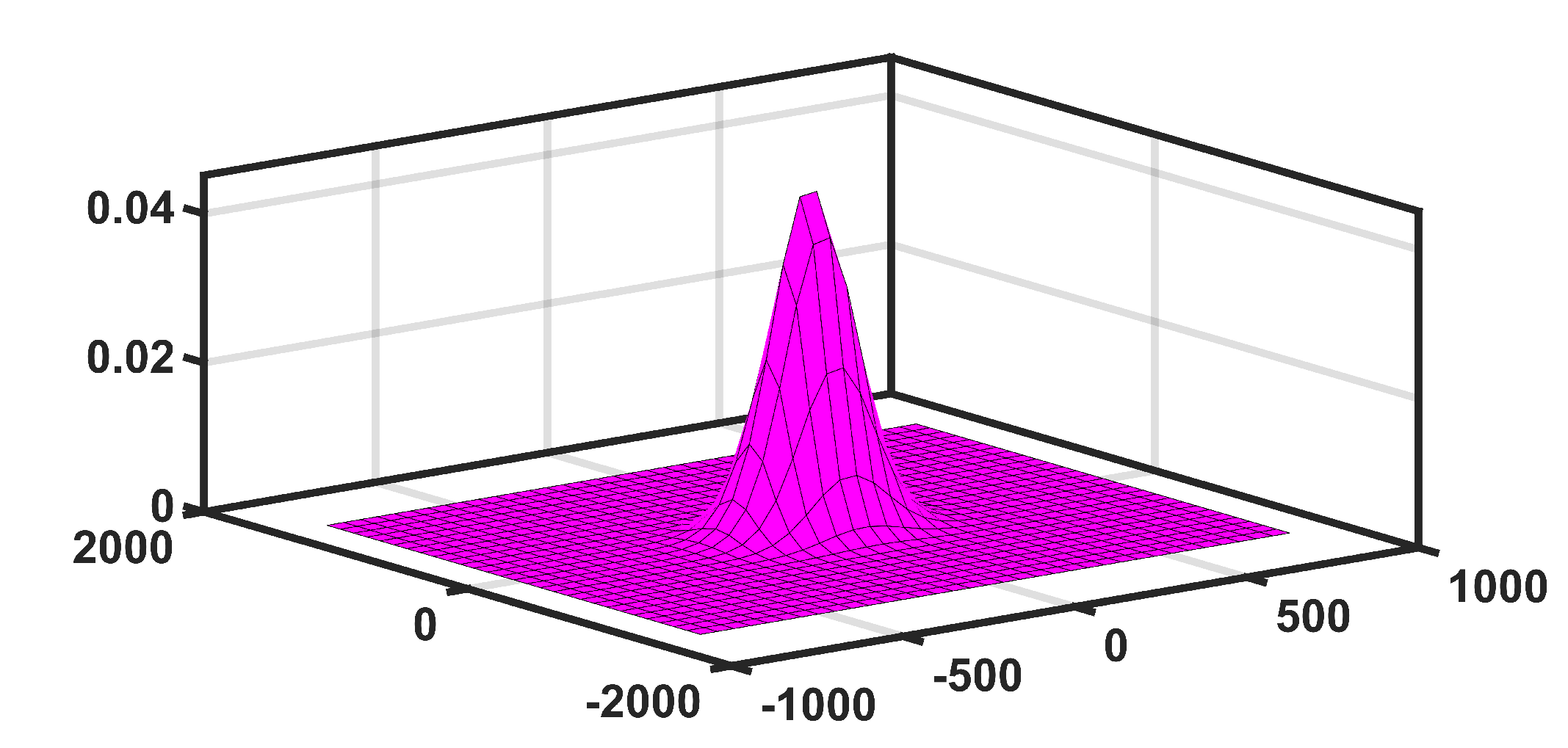}
          \caption{}
          \label{mpdf_isom_Gamma}
      \end{subfigure}
      \begin{subfigure}{0.31\textwidth}
        \includegraphics[width=\textwidth]{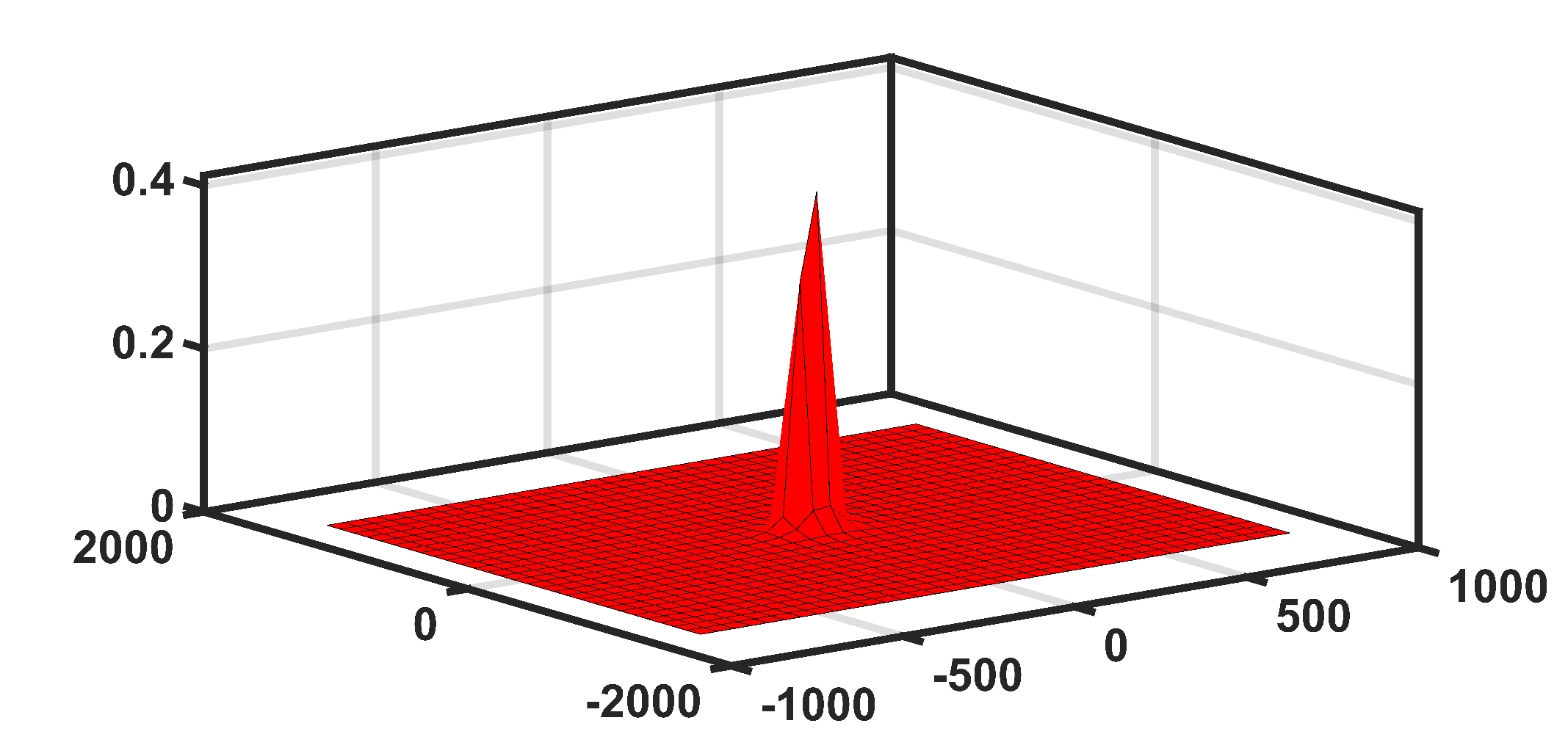}
          \caption{}
          \label{mpdf_isom_SMM}
      \end{subfigure}
      
     \caption {Visual comparisons between (a) empdf (yellow) and estimated pdf's from models: (b) CG-E (blue), (c) CG-G (magenta) and (d) CG-IG (red) for isometric activity during $6$ kg lifting corresponding to the subject-5 and trial-8.}
      \label{fig:VisualComp1}
\end{figure*}

\begin{figure*}[h]
\captionsetup[subfigure]{justification=centering}
    \centering
      \begin{subfigure}{0.31\textwidth}
        \includegraphics[width=\textwidth]{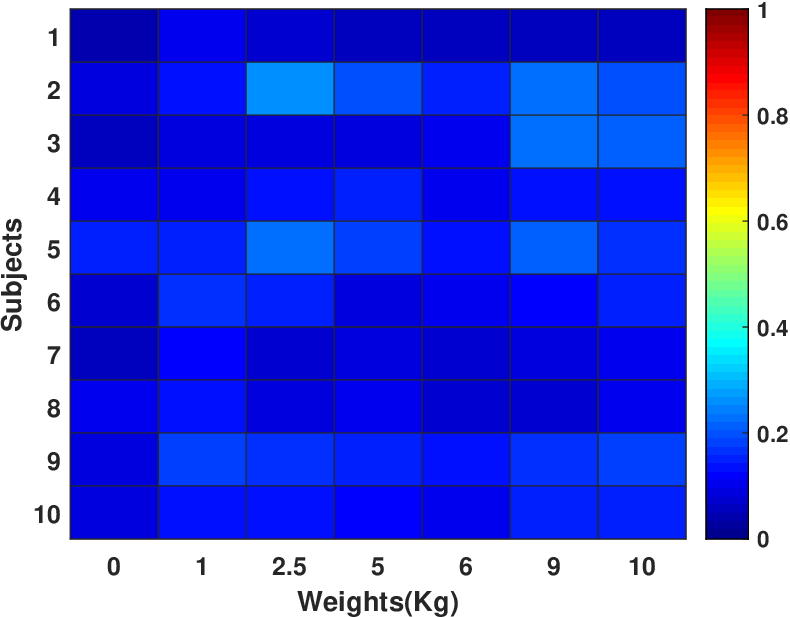} 
          \caption{}
          \label{fig:NiceImage1}
      \end{subfigure}
      \begin{subfigure}{0.31\textwidth}
        \includegraphics[width=\textwidth]{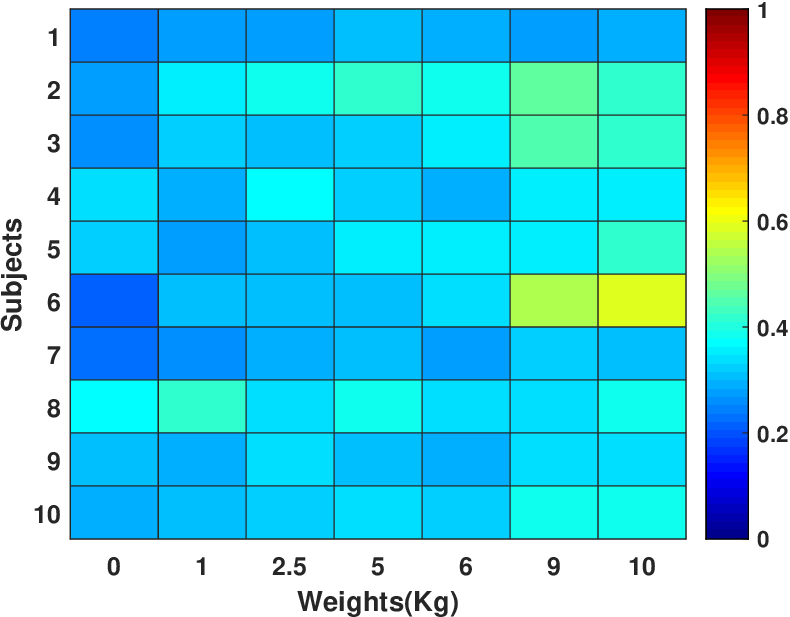}
          \caption{}
          \label{fig:NiceImage2}
      \end{subfigure}
      \begin{subfigure}{0.31\textwidth}
        \includegraphics[width=\textwidth]{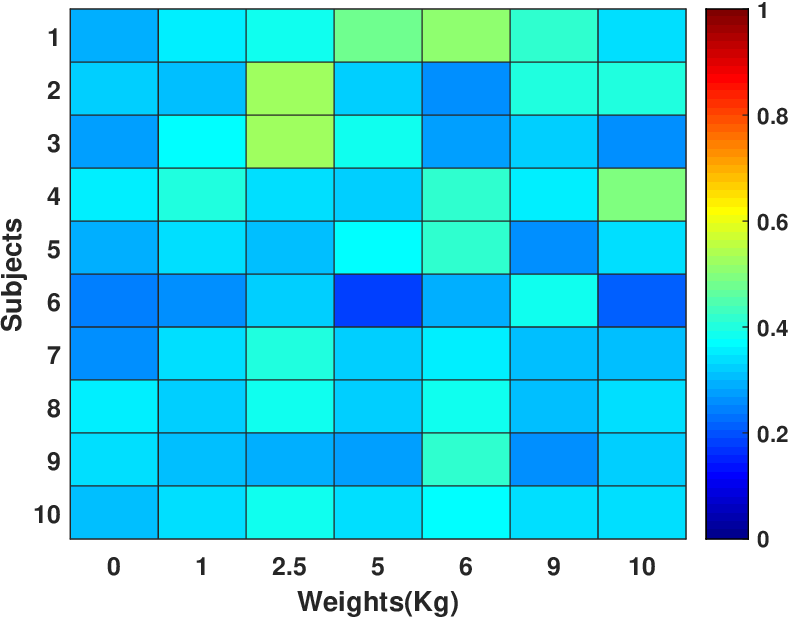} 
          \caption{}
          \label{fig:NiceImage3}
      \end{subfigure}
      \begin{subfigure}{0.31\textwidth}
        \includegraphics[width=\textwidth]{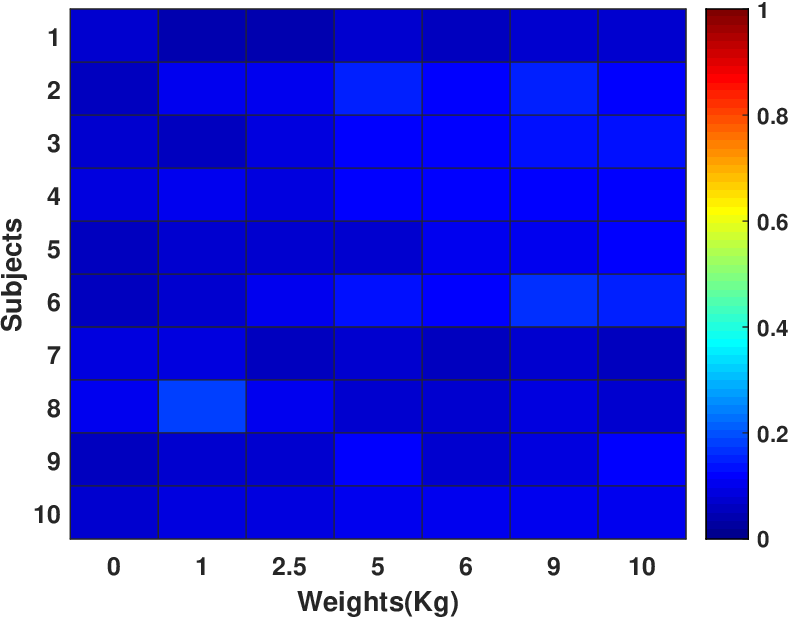}
          \caption{}
          \label{fig:NiceImage1}
      \end{subfigure}
      \begin{subfigure}{0.31\textwidth}
        \includegraphics[width=\textwidth]{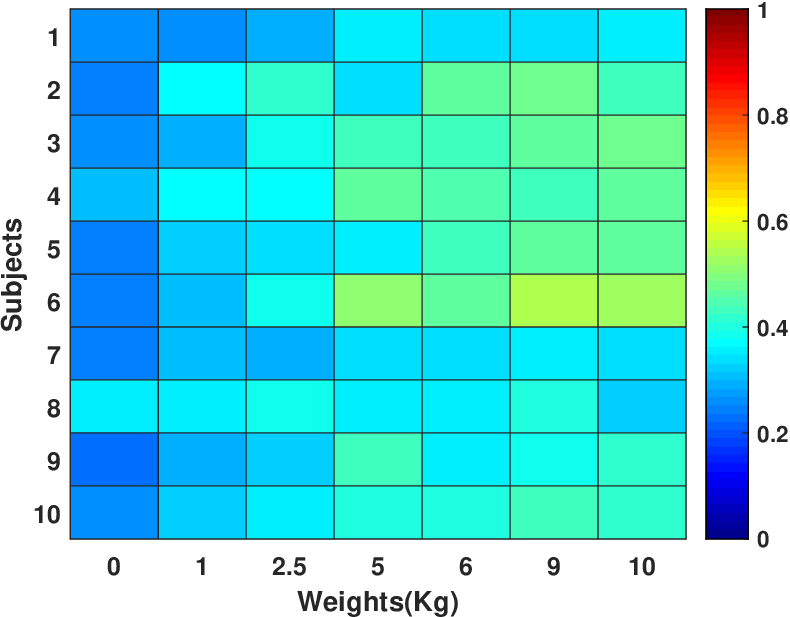}
          \caption{}
          \label{fig:NiceImage2}
      \end{subfigure}
      \begin{subfigure}{0.31\textwidth}
        \includegraphics[width=\textwidth]{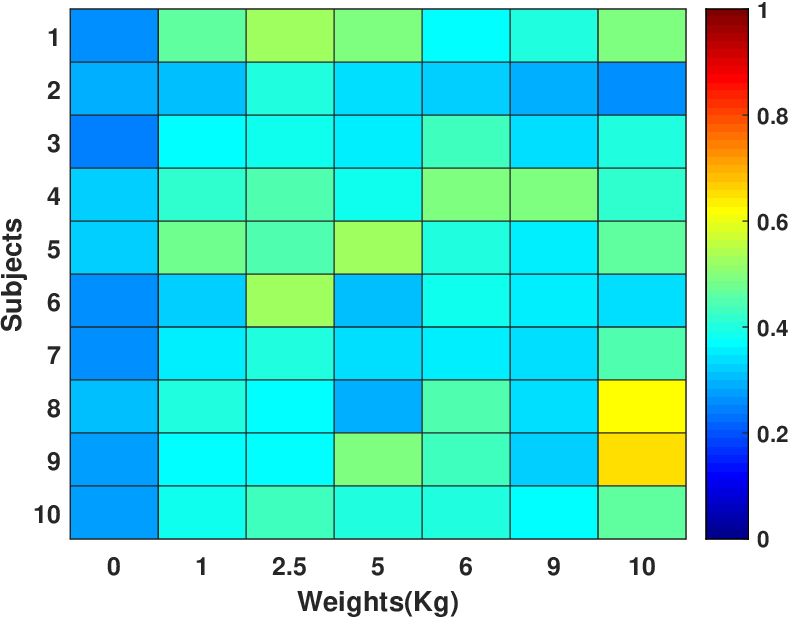}
          \caption{}
          \label{fig:NiceImage3}
      \end{subfigure}
      \caption{Heatmaps of KLD : (a) CG-E, (b) CG-G and (c) CG-IG corresponding to experiment-I, (d) CG-E, (e)  CG-G and (f) CG-IG from experiment-II}
      \label{heatmap}
\end{figure*}

\begin{figure*}[t]
\captionsetup[subfigure]{justification=centering}
    \centering
      \begin{subfigure}{0.45\textwidth}        \includegraphics[width=\textwidth]{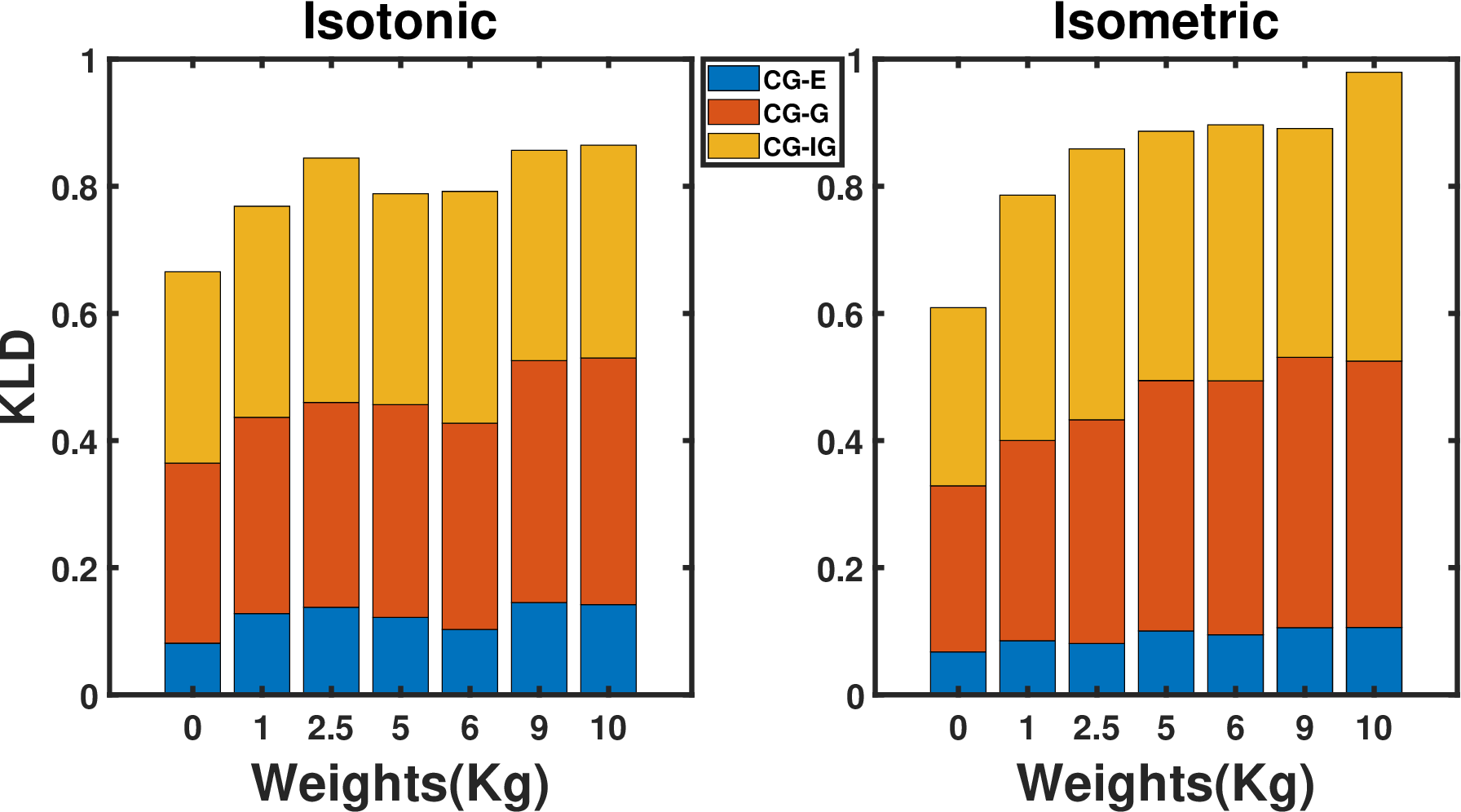} 
          \caption{} 
        \label{Avg_Sub_isot_isom}
      \end{subfigure}
      \begin{subfigure}{0.45\textwidth}
        \includegraphics[width=\textwidth]{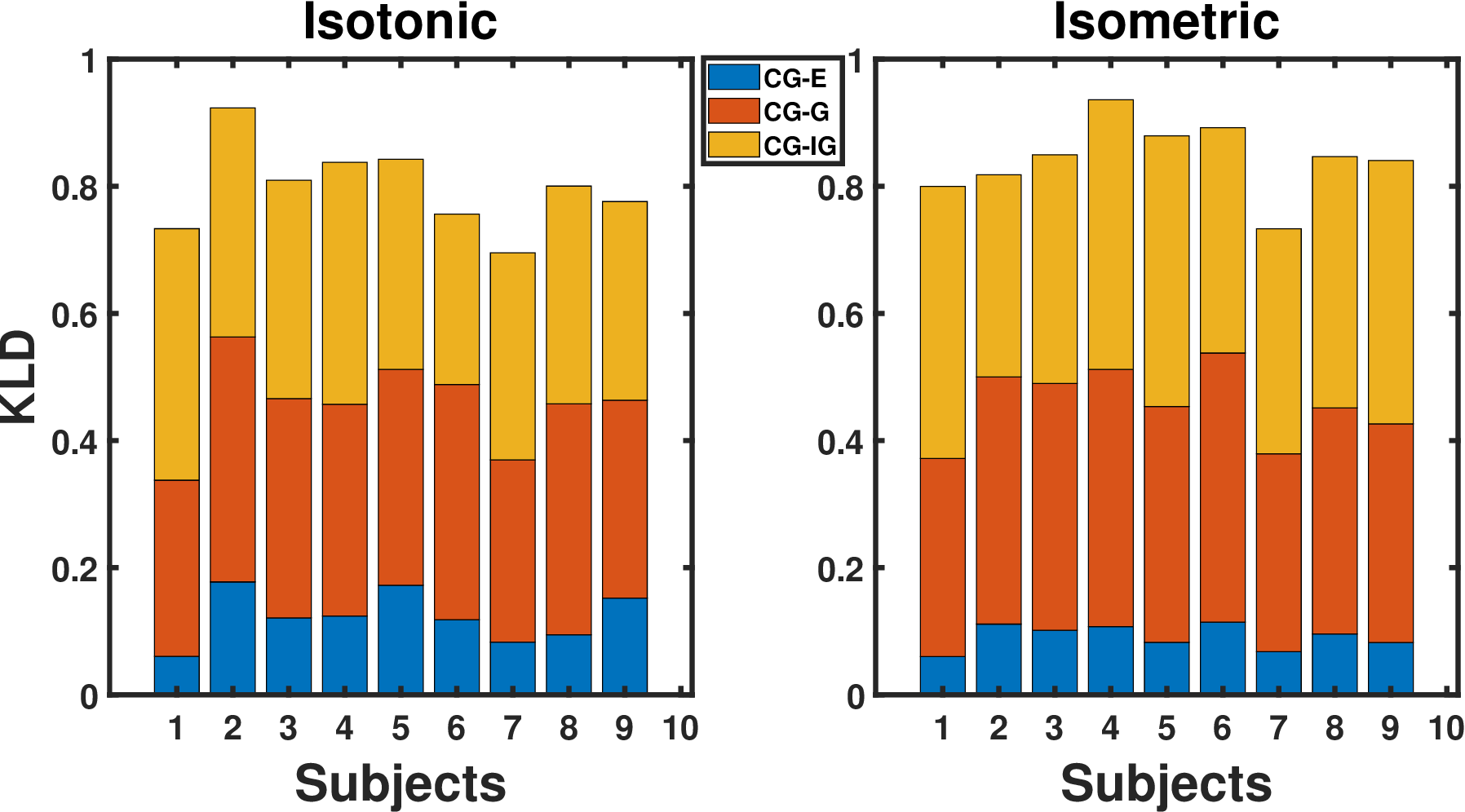}
          \caption{}
        \label{Avg_Wei_isot_isom}
      \end{subfigure}
  \caption{KLD of CG-E, CG-G and CG-IG for isometric and isotonic activities: (a) averaged across subjects and plot vs weights (b) averaged across weights and plot vs subjects}
     \label{KLD_1D_plots}
\end{figure*}

\section{Model Analysis and Discussion} \label{sec:Results}

In this section, the most suitable model for the EMAHA-DB2 data is determined by comparing the following compound Gaussian models:  
\begin{itemize}
    \item CG-E (Proposed model)
    \item CG-IG \cite{furui2019scale,furui2021emg} 
    \item CG-G  \cite{wang2006maximum}
\end{itemize}
Model validation is carried out for each of the sEMG signals corresponding to the experiments in section-III using the  following evaluation methods. 
\begin{itemize}
    \item  Qualitative analysis based on visual inspection
    \item  Quantitative analyses: 
    \begin{enumerate}
    \item Moment analysis
    \item  Analysis of KLD
    \item  Coefficient of determination (COD) R-squared 
    \item  Log-likelihood values
    \end{enumerate}
\end{itemize}

\subsection{{Visual Inspection}} 

The Figs. \ref{fig:VisualComp} and \ref{fig:VisualComp1} illustrates the empdf (yellow) and the models from CG-E (blue), CG-G (magenta) and CG-IG (red) estimated for the strength of two channel sEMG signals. 
Specifically, Fig. \ref{fig:VisualComp} illustrates the results from analysis on sEMG signals of experiment I (isotonic activity) corresponds to subject-1 while training with $6$kg dumbbell.  
and Fig. \ref{fig:VisualComp1} corresponds to the experiment II (isometric activity) with $6$kg dumbbell. 
From these figures, it is noticed that the CG-E model fits the empdf better in comparison to other models. The models CG-G and CG-IG are weaker fits compared to CG-E. A similar analysis is carried out for the rest of the data and  it is observed that CG-E model has the best agreement with the empdf among the three compound models. 

\begin{table*}[htb]
\centering
\caption{Estimated moments of isotonic activity during 6 kg lifting corresponding to the subject-1 and trial-8 } 
\begin{tabular}{cclll}
\hline
\textbf{Estimates } & \textbf{empdf}                                                                                         & ~~~~~~~~\textbf{CG-E}                                                                                       &
~~~~~~~~\textbf{ CG-IG  }                                                                            & 
~~~~~~~~\textbf{CG-G}                                                                                     \\ \hline
Mean       & 
$\begin{bmatrix}
  -8.2495 & 0
\end{bmatrix}*10^{-8}$                                                                          & $\begin{bmatrix}
-0.0006551  & -0.0003244 
\end{bmatrix}$                                                                      & $\begin{bmatrix}
0.0446   & 0.0155
\end{bmatrix}$                                                         & $\begin{bmatrix}
-0.001786  & -0.001169

\end{bmatrix} $                                                                \\ \hline
Covariance & $\begin{bmatrix}
3.1235  &  0.0101 \\
    0.0101  &  1.1892
\end{bmatrix}* 10^4$ & $\begin{bmatrix}
3.3269   & 0.0535 \\
    0.0535   & 1.0009
\end{bmatrix}* 10^4 $& $\begin{bmatrix}
93.2162  &  1.5157\\
    1.5157 &  27.3331
\end{bmatrix}$& $\begin{bmatrix}
2.2445 &   0.0272 \\
    0.0272  &  0.7818
\end{bmatrix}* 10^4$\\ \hline
Mardia's Kurtosis   & 
8.3029                                                                   &     ~~~~~~~~~~7.7038                                                                   
& ~~~~~~~~~~13.6497                                                            & ~~~~~~~~~~7.1717  
 \\ \hline
\end{tabular}

 \label{Est_isotonic}
\end{table*}

\begin{table*}[t!]
\centering
\caption{Estimated moments of isometric activity during 6 kg lifting corresponding to the subject-5 and trial-8} 
\begin{tabular}{cclll}
\hline
\textbf{Estimates } & \textbf{empdf}                                                                                         & ~~~~~~~~\textbf{CG-E}                                                                                       &
~~~~~~~~\textbf{ CG-IG  }                                                                            & 
~~~~~~~~\textbf{CG-G}                                                                                     \\ \hline
Mean       & $\begin{bmatrix}
 1.2140 & -4.5297
\end{bmatrix}*10^{-8}$                                                                         & $\begin{bmatrix}
 0.000485  & -0.001479 
\end{bmatrix}$                                                                    & 
$\begin{bmatrix}
0.0089  & -0.0078
\end{bmatrix} $                                                        & $\begin{bmatrix}
0.000800   -0.002332
\end{bmatrix} $                                                                 \\ \hline
Covariance & $\begin{bmatrix}
3.7800 &  -0.1106\\
   -0.1106 &   7.6191
\end{bmatrix}*10^4$
& 
$\begin{bmatrix}
3.7561  & -0.0836 \\
   -0.0836  &  6.6507
\end{bmatrix}*10^4 $& $\begin{bmatrix}
 146.9145  & -3.3232 \\
   -3.3232 & 261.9663
\end{bmatrix}$& $\begin{bmatrix}
4.2396  & -0.0794\\
   -0.0794  &  5.0369
\end{bmatrix}*10^4$\\ \hline
Mardia's Kurtosis    & 
8.7624
& 
    ~~~~~~~~~~7.9451                                                                  & 
~~~~~~~~~~13.0776                                                    & 
~~~~~~~~~~7.6927
                                    \\ \hline
      
\end{tabular}
 \label{Est_isometric}
\end{table*}

\begin{table*}[t!]
\centering
\caption{Averaged estimated moments across all subjects and trials related to isotonic activity } 
\begin{tabular}{cclll}
\hline
\textbf{Estimates } & \textbf{empdf}                                                                                         & ~~~~~~~~\textbf{CG-E}                                                                                       &
~~~~~~~~\textbf{ CG-IG  }                                                                            & 
~~~~~~~~\textbf{CG-G}                                                                                     \\ \hline
Mean       & $\begin{bmatrix}
0 & 0
\end{bmatrix} $                                                                         & $\begin{bmatrix}
0.00091  & 0.00035 
\end{bmatrix} $                                                                    & $\begin{bmatrix}
0.00620  & 0.00087
\end{bmatrix} $                                                        & $\begin{bmatrix}
0.00284  & 0.00075
\end{bmatrix} $                                                                  \\ \hline
Covariance & $\begin{bmatrix}
3.0096 &  0.0652\\
   0.0652 &   1.3743
\end{bmatrix}*10^4$ & 
$\begin{bmatrix}
3.0271  & 0.0708\\
   0.0708  &  1.2168
\end{bmatrix}*10^4$ & $\begin{bmatrix}
 122.160  & 2.8723\\
  2.8723 & 48.2928
\end{bmatrix}$& $\begin{bmatrix}
3.8457  & 0.0502 \\
   0.0502  & 1.9896
\end{bmatrix}*10^4$\\ \hline
Mardia's Kurtosis    & 
7.2156
& 
    ~~~~~~~~~~6.5821                                                                  & 
~~~~~~~~~~14.1648                                                    & 
~~~~~~~~~~6.2847
                                    \\ \hline
      
\end{tabular}
 \label{Avg_Est_isotonic}
\end{table*}

\subsection{Quantitative Analysis}
\subsubsection{Moment Analysis}
{The estimated moments such as the mean, covariance and the Mardia’s kurtosis\cite{mardia1974applications,cain2017univariate} corresponding to Fig.\ref{fig:VisualComp} and \ref{fig:VisualComp1} are shown in table-\ref{Est_isotonic} and \ref{Est_isometric}. Among the three models, the moments of CG-E are best match to those of the empdf. In addition, the averaged moments across the subjects and trials for the isotonic activity corresponding to $6$kg weight lifting are presented in table \ref{Avg_Est_isotonic}. These results indicate agreement between the moments corresponding to the CG-E and the empdf.
} 

\subsubsection{KL-divergence}
\label{KLD-SECTION}
\begin{figure*}[h!]
\captionsetup[subfigure]{justification=centering}
    \centering
      \begin{subfigure}{0.45\textwidth}
        \includegraphics[width=\textwidth]{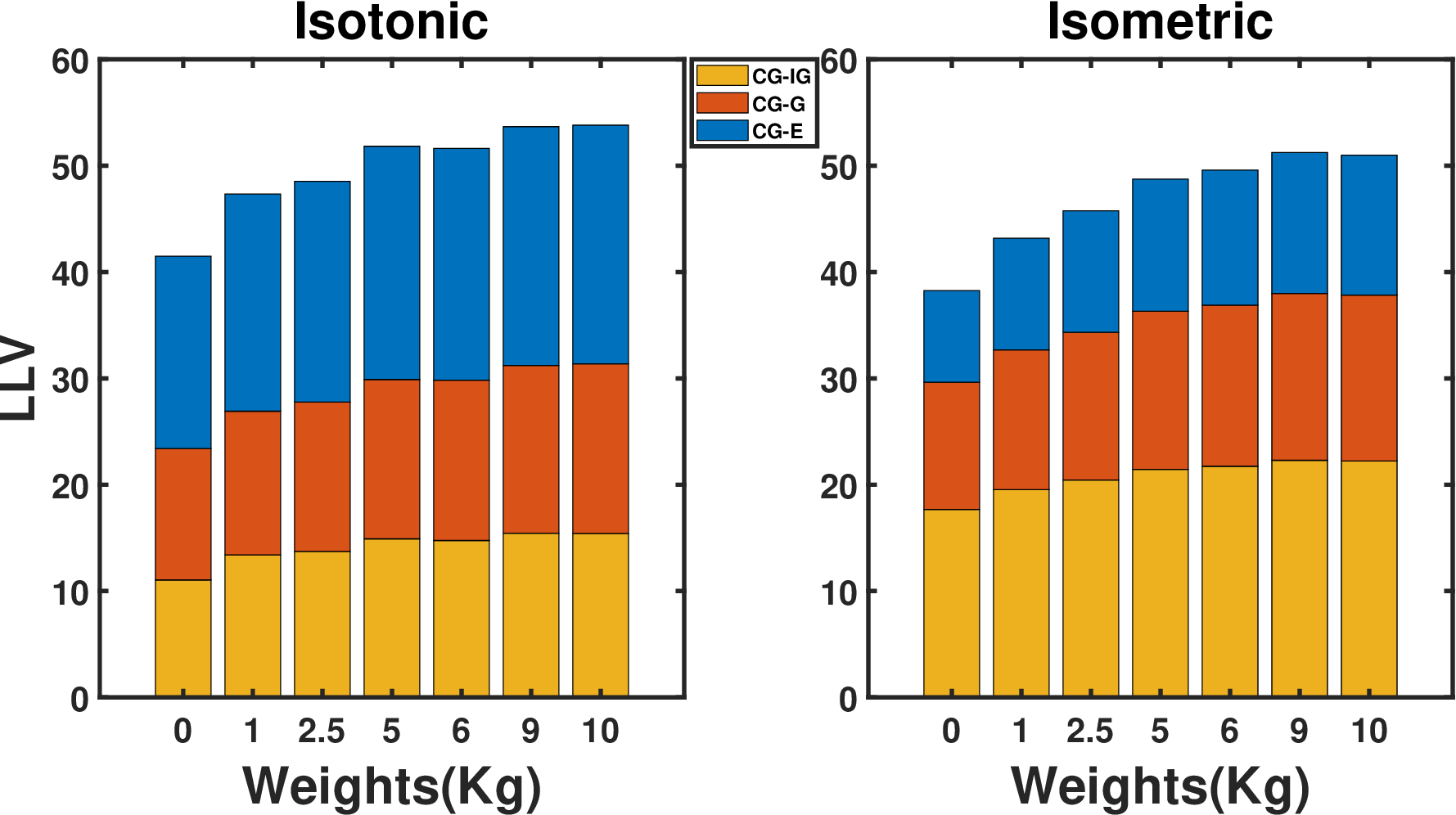} 
          \caption{}
\label{Avg_Likeli_Sub_isot_isom}
      \end{subfigure}
      \begin{subfigure}{0.45\textwidth}
        \includegraphics[width=\textwidth]{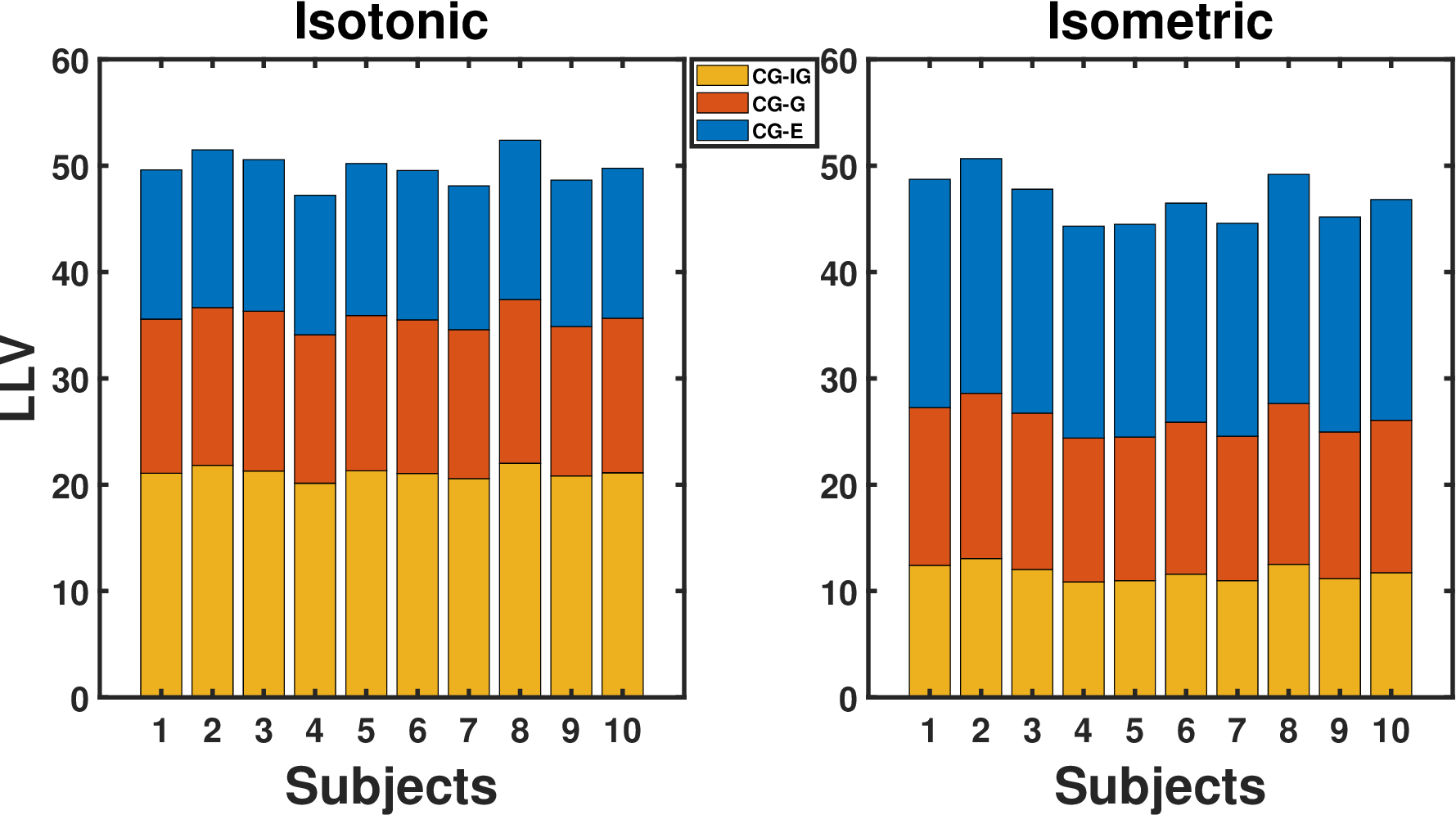}
          \caption{}
\label{Avg_Likeli_Wei_isot_isom}
      \end{subfigure}
  \caption{LLV of CG-E, CG-G and CG-IG for isometric and isotonic activities: (a) averaged across subjects and plot vs weights (b) averaged across weights and plot vs subjects}
\label{Likelihood}
\end{figure*}

\begin{figure*}[t]
\captionsetup[subfigure]{justification=centering}
    \centering
      \begin{subfigure}{0.46\textwidth}
        \includegraphics[width=\textwidth]{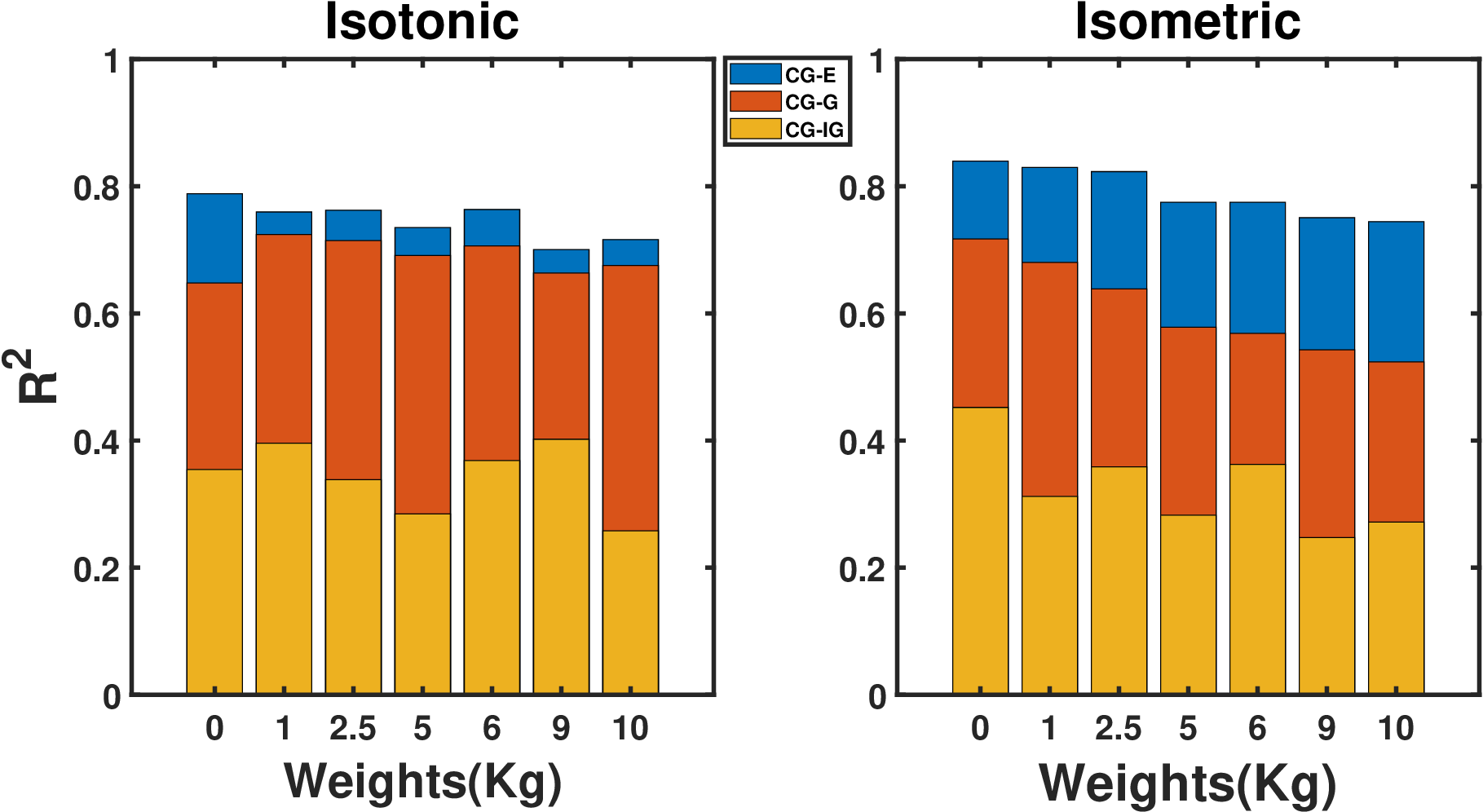} 
          \caption{}
\label{Avg_Sub_isot_isom_R2_v2}
      \end{subfigure}
      \begin{subfigure}{0.45\textwidth}
        \includegraphics[width=\textwidth]{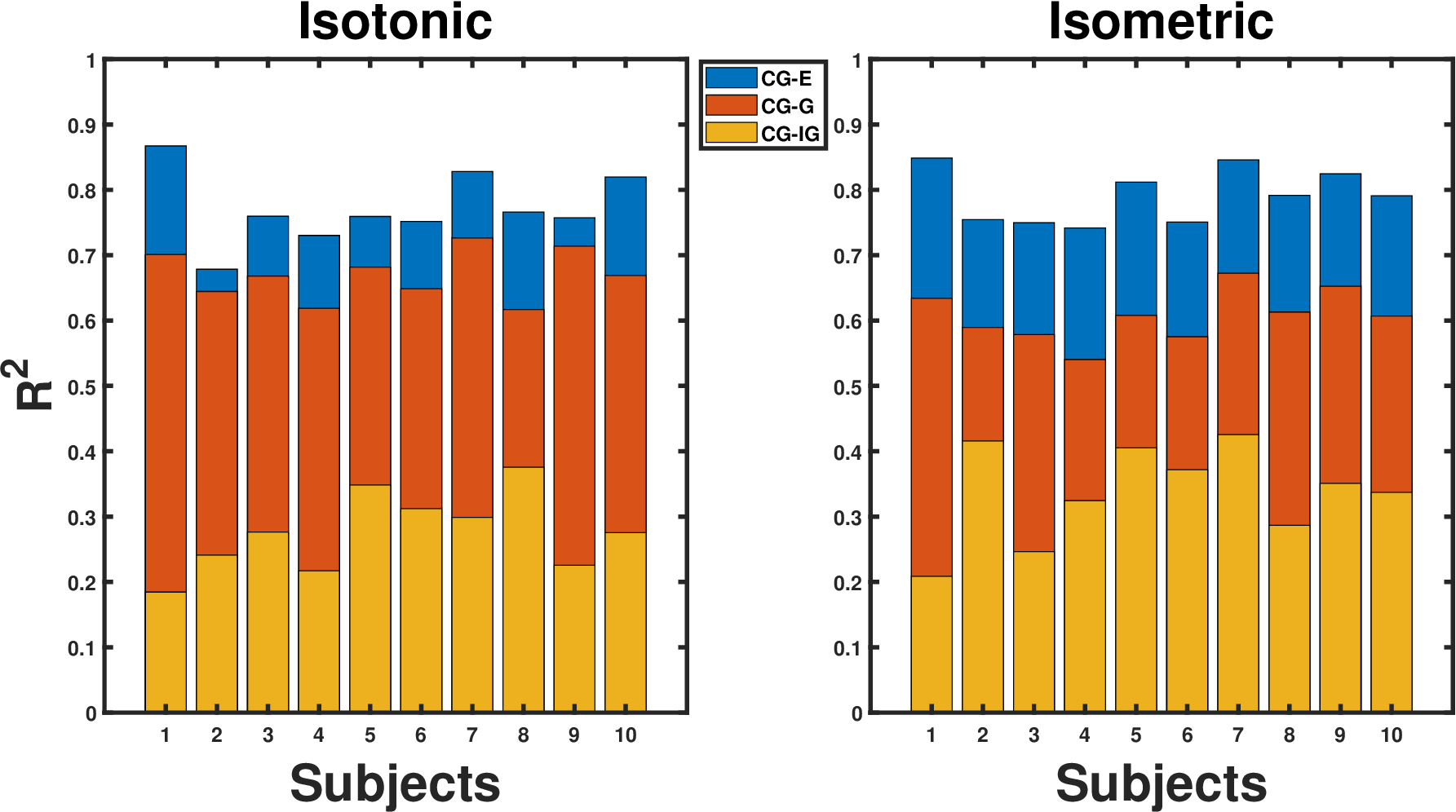}
          \caption{}
\label{Avg_Wei_isot_isom_R2_v2}
      \end{subfigure}
  \caption{$R^2$ of CG-E, CG-G and CG-IG for isometric and isotonic activities: (a) averaged across subjects and plot vs weights (b) averaged across weights and plot vs subjects}
\label{R-squared}
\end{figure*}
In this study, the KLD is evaluated between the empdf and the three compound models and shown in  Figs. \ref{heatmap} to \ref{KLD_1D_plots}. Fig. \ref{heatmap}, illustrates KLD heatmaps as a function of subjects and activities. The KLD in each cell of the heatmap is an average over the trials of the corresponding activity. The KLDs corresponding to experiment-I are shown in Figs. \ref{heatmap} (a) to (c), while the KLDs corresponding to experiment-II are presented in Figs. \ref{heatmap} (d) to (f). Based on these heatmaps, it is observed that the CG-E model has the lowest KLD among the three compound models. The ranges of KLD for the heatmaps in Fig. \ref{heatmap} are given in table \ref{tab:kldBound}. For experiment II, the maximum KLD from the CG-E does not exceed the minimum KLD from the CG-G and CG-IG. In the case of experiment I, the maximum KLD from the CG-E is less than half the maximum from the other models.

Fig. \ref{Avg_Sub_isot_isom}  shows the averaged KLD across the subjects as a function of the weights and Fig. \ref{Avg_Wei_isot_isom} shows the vice-versa. The KLD of the CG-E, CG-G and CG-IG are represented in blue, orange and yellow  respectively.  From the Figs. \ref{Avg_Sub_isot_isom} and \ref{Avg_Wei_isot_isom}, it is noted that for both the experiments, the averaged KLD corresponding to either the subjects or the weights is the lowest for the CG-E, when compared to CG-G and CG-IG. {As mentioned earlier, the optimal choice $(K,N) = (200,80)$ is made based on grid search for lowest KLD over a region of possible values for $K$  and $N$.}

\subsubsection{Log-Likelihood values}

Fig. \ref{Avg_Likeli_Sub_isot_isom} illustrates the LLV averaged across the subjects and trials as a function of the weights.  Fig. \ref{Avg_Likeli_Wei_isot_isom} shows the LLV averaged across the weights and trials as a function of the subjects. The LLV of the CG-E, CG-G and CG-IG are shown again in blue, orange and yellow respectively. From these figures it can be observed that for both the experiments, the averaged LLV, for both the  subjects and the weights, is the highest for the CG-E among the three models.

\subsubsection{Coefficient of determination (COD) $R^2$}

The averaged $R^2$ for CG-E, CG-G and CG-IG corresponding to  experiments-I and II are illustrated in Fig \ref{R-squared}. Specifically, Fig \ref{Avg_Sub_isot_isom_R2_v2} shows $R^2$ averaged across subjects vs. weights and Fig. \ref{Avg_Wei_isot_isom_R2_v2} shows vice-versa. From these figures it is obvious that  $R^2$ associated with the CG-E is the highest among the models and followed by that of the CG-G and the CG-IG. It is also evident that the difference in $R^{2}$ between CG-E and CG-G models is lesser in the experiment-I however it is much greater with respect to experiment-II. The minimum and maximum values of $R^2$ as functions of subjects and weights for each activity are shown in tables-\ref{tab:S_R2} and \ref{tab:W_R2} respectively.

 \begin{table}[h] 
 \centering
 \caption{Minimum and maximum values of KLD}
\begin{tabular}{cccc}
\hline
\multicolumn{1}{l}{Experiment}                  & CG-E                  & CG-G            & CG-IG            \\ \hline
I                            & {[}0.0402  0.2506{]} & {[}0.214   0.5878{]} & {[}0.1817 0.5225{]}  \\ 
II                             & {[}0.04487  0.1721{]} & {[}0.2286    0.5335{]} & {[}0.2428 0.65{]} \\ \hline

\end{tabular} \label{tab:kldBound}
\end{table}

 \begin{table}[h] 
 \centering
 \caption{Minimum and maximum  $R^2$ averaged across subjects }
\begin{tabular}{cccc}
\hline
\multicolumn{1}{l}{Experiment}                  & CG-E                  & CG-G            & CG-IG             \\ \hline
I                             & {[}0.7005  0.7882{]} & {[}0.6478  0.7243{]} & {[}0.2580  0.4021{]}  \\ 
II                             & {[}0.7444  0.8395{]} & {[}0.5237   0.7171{]} & {[}0.2474  0.4521{]} \\ \hline
\end{tabular}  \label{tab:S_R2}
\end{table}

 \begin{table}[h] 
 \centering
 \caption{Minimum and maximum  $R^2$ averaged across weights }
\begin{tabular}{cccc}
\hline
\multicolumn{1}{l}{Experiment}                  & CG-E                  & CG-G            & CG-IG             \\ \hline
I                             & {[}0.6488  0.8672{]} & {[}0.6169   0.7593{]} & {[}0.1845  0.3754{]}  \\ 
II                             & {[}0.7417  0.8487{]} & {[}0.5404  0.6723{]} & {[}0.2087  0.4254{]} \\ \hline
\end{tabular}  \label{tab:W_R2}
\end{table}

\subsection{Analysis of rate parameter ($\lambda$) of CG-E model}
\begin{figure}[t]
        \centering
      \includegraphics[width=0.91\columnwidth]{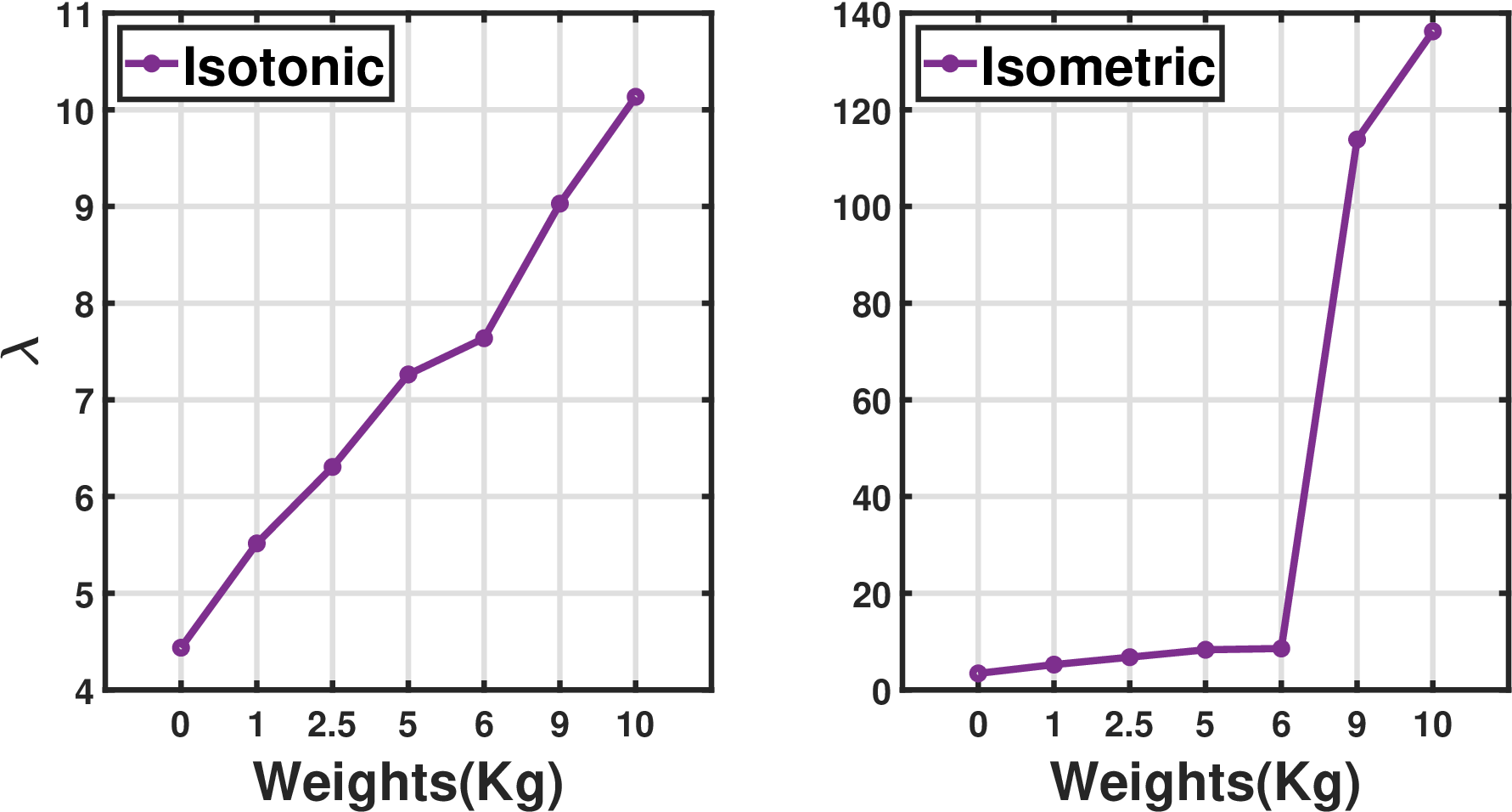}
      \caption{Estimate of rate parameter ($\lambda$) as a function of weights averaged across trials and subjects}
     \label{R_P}
\end{figure}

 Fig. \ref{R_P} depicts the estimates of the rate parameter ($\lambda$) that corresponds to the CG-E as a function of dumbbell weights. Recall that $\lambda$ originally quantifies the statistical mean of the texture $z_k$. 
The estimate of $\lambda$ shown in Fig. \ref{R_P} is an average over the trials and subjects. Figures on the left and right correspond to experiments I  and II respectively. First, it is clear that the value of $\hat{\lambda}$ increases with the dumbbell weight. Additionally, note that the variation of $\hat{\lambda}$ is less with the experiment $I$ (isotonic) and more with the experiment II (isometric). 
Furthermore in the experiment II, the esimtate $\hat{\lambda}$ doesnot increase significantly till $6$ kgs lifiting weight. However, it rises rapidly at higher weights $9,10$ kgs. Note that the forces corresponding to isometrics are stronger that those in the isotonic contractions \cite{reed2008principles}. The amount of muscle force (or) muscles recruitment required for a weight lift is generally proportional to its weight. Thus from Fig. \ref{R_P}, muscle force required to lift a weight can be attributed to the rate parameter ($\lambda$). Note that the motor units recruitment may increase with the force generated. Thus for strength training athletes, the muscle force and rate of muscle force generated can be correlated to the texture variable's estimated mean $\hat{\lambda}$. 

\begin{figure}[h]
        \centering
      \includegraphics[width=0.9\columnwidth]{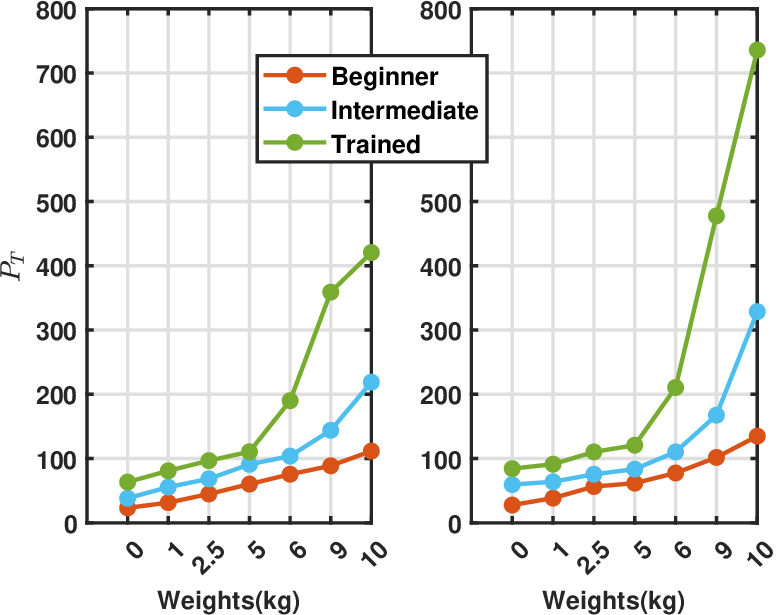}
      \caption{Sum of estimated variances $P_{T}$ as a function of weights averaged across trials and subjects}
     \label{based_experience}
\end{figure}
\subsection{Analysis of sum of variances}
The metric $P_{T}$
denotes the square root of sum of variances (trace of $\Sigma$) from
BB and FCU.
\begin{equation}
    P_T=\sqrt{\sigma^{2}_{T,BB}+\sigma^{2}_{T,FCU}}
\end{equation}
{where $\sigma^{2}_{T,BB}$, $\sigma^{2}_{T,FCU}$ are the variances corresponding to BB and FCU.} The $P_{T}^{2}$
termed as \textit{T-power} is sum of variances of the sEMG signal from BB and FCU and can be related to the
muscle force. Fig. \ref{based_experience} depicts $P_T$ from BB and FCU for
isotonic (left) and isometric (right) activities. From this figure, it is interesting to note that both for isotonic and isometric activities, for any lifting load, the total signal power seems to be directly related to the subject’s experience. For any weight, the trained subjects produced the highest T-power while the beginners generated the least T-power. Additionally, $P_T$ can be correlated with a subject’s strength. A higher slope for $P_T$ vs weights indicates higher strength to lift heavier weights. Based on Fig. \ref{based_experience}, the trained group has a steeper slope of $P_T$, followed by the intermediate and the beginner groups.

\section{Conclusion and Future work}
In this paper, an multivariate Compound-Gaussian model is proposed for sEMG signals by considering improved variance as exponential model. This model is compared with existing (CG-IG) model. In addition to this, the proposed model is also compared with the CG model in which variance is modeled as gamma. The goodness of the model is justified using, (1) A qualitative comparison with the empdf reveals the best agreement with the CG-E model, (2) the KLD between the fitted model and the empdf, again the KLD is lowest for the CG-E model, (3) Coefficient of determination (COD) - $R^2$  here it is noted that $R^2$ in case of the CG-E model is closest to unity and (4) the Log-Likelihood values (LLV) that also support the CG-E model. Finally, the estimtates of the rate parameter ($\lambda$) and the signal covariance of the (CG-E) model is analyzed in different measurement conditions. In future work, the plans include statistical modeling of sEMG signals corresponding to sports activities and understand their role in muscle coordination.

\bibliographystyle{IEEEtran}
\bibliography{References.bib}

\begin{thebibliography}{10}
\providecommand{\url}[1]{#1}
\csname url@samestyle\endcsname
\providecommand{\newblock}{\relax}
\providecommand{\bibinfo}[2]{#2}
\providecommand{\BIBentrySTDinterwordspacing}{\spaceskip=0pt\relax}
\providecommand{\BIBentryALTinterwordstretchfactor}{4}
\providecommand{\BIBentryALTinterwordspacing}{\spaceskip=\fontdimen2\font plus
\BIBentryALTinterwordstretchfactor\fontdimen3\font minus
  \fontdimen4\font\relax}
\providecommand{\BIBforeignlanguage}[2]{{%
\expandafter\ifx\csname l@#1\endcsname\relax
\typeout{** WARNING: IEEEtran.bst: No hyphenation pattern has been}%
\typeout{** loaded for the language `#1'. Using the pattern for}%
\typeout{** the default language instead.}%
\else
\language=\csname l@#1\endcsname
\fi
#2}}
\providecommand{\BIBdecl}{\relax}
\BIBdecl

\bibitem{wang2006simulation}
W.~Wang, A.~D. Stefano, and R.~Allen, ``A simulation model of the surface {EMG}
  signal for analysis of muscle activity during the gait cycle,''
  \emph{Computers in biology and medicine}, vol.~36, no.~6, pp. 601--618, 2006.

\bibitem{stashuk1993simulation}
D.~W. Stashuk, ``Simulation of electromyographic signals,'' \emph{Journal of
  Electromyography and Kinesiology}, vol.~3, no.~3, pp. 157--173, 1993.

\bibitem{cuddon2002electrophysiology}
P.~A. Cuddon, ``Electrophysiology in neuromuscular disease,'' \emph{Veterinary
  Clinics: Small Animal Practice}, vol.~32, no.~1, pp. 31--62, 2002.

\bibitem{fleischer2006application}
C.~Fleischer, A.~Wege, K.~Kondak, and G.~Hommel, ``Application of emg signals
  for controlling exoskeleton robots,'' 2006.

\bibitem{clancy1999probability}
E.~A. Clancy and N.~Hogan, ``Probability density of the surface electromyogram
  and its relation to amplitude detectors,'' \emph{IEEE Transactions on
  Biomedical Engineering}, vol.~46, no.~6, pp. 730--739, 1999.

\bibitem{vigotsky2018interpreting}
A.~D. Vigotsky, I.~Halperin, G.~J. Lehman, G.~S. Trajano, and T.~M. Vieira,
  ``Interpreting signal amplitudes in surface electromyography studies in sport
  and rehabilitation sciences,'' \emph{Frontiers in physiology}, vol.~8, p.
  985, 2018.

\bibitem{de1997use}
C.~J. De~Luca, ``The use of surface electromyography in biomechanics,''
  \emph{Journal of applied biomechanics}, vol.~13, no.~2, pp. 135--163, 1997.

\bibitem{soderberg1984electromyography}
G.~L. Soderberg and T.~M. Cook, ``Electromyography in biomechanics,''
  \emph{Physical Therapy}, vol.~64, no.~12, pp. 1813--1820, 1984.

\bibitem{hasanbelliu2004multi}
E.~Hasanbelliu, ``A multi-dimensional visualization tool for understanding the
  role of {EMG} signals in head movement anticipation,'' in \emph{ACM SIGGRAPH
  2004 Posters}, 2004, p. 109.

\bibitem{furui2019scale}
A.~Furui, H.~Hayashi, and T.~Tsuji, ``A scale mixture-based stochastic model of
  surface {EMG} signals with variable variances,'' \emph{IEEE Transactions on
  Biomedical Engineering}, vol.~66, no.~10, pp. 2780--2788, 2019.

\bibitem{parker1977signal}
P.~A. Parker, J.~A. Stuller, and R.~N. Scott, ``Signal processing for the
  multistate myoelectric channel,'' \emph{Proceedings of the IEEE}, vol.~65,
  no.~5, pp. 662--674, 1977.

\bibitem{hogan1980myoelectric}
N.~Hogan and R.~W. Mann, ``Myoelectric signal processing: Optimal estimation
  applied to electromyography-{Part I}: Derivation of the optimal
  myoprocessor,'' \emph{IEEE Transactions on Biomedical Engineering}, no.~7,
  pp. 382--395, 1980.

\bibitem{der1998detection}
V.~Der~Bilt and V.~Der~Glas, ``Detection of onset and termination of muscle
  activity in surface electromyograms,'' \emph{Journal of oral rehabilitation},
  vol.~25, no.~5, pp. 365--369, 1998.

\bibitem{milner1975relation}
H.~Milner-Brown and R.~Stein, ``The relation between the surface electromyogram
  and muscular force.'' \emph{The Journal of physiology}, vol. 246, no.~3, pp.
  549--569, 1975.

\bibitem{hunter1987estimation}
I.~Hunter, R.~Kearney, and L.~Jones, ``Estimation of the conduction velocity of
  muscle action potentials using phase and impulse response function
  techniques,'' \emph{Medical and Biological Engineering and Computing},
  vol.~25, no.~2, pp. 121--126, 1987.

\bibitem{bilodeau1997normality}
M.~Bilodeau, M.~Cincera, A.~B. Arsenault, and D.~Gravel, ``Normality and
  stationarity of {EMG} signals of elbow flexor muscles during ramp and step
  isometric contractions,'' \emph{Journal of Electromyography and Kinesiology},
  vol.~7, no.~2, pp. 87--96, 1997.

\bibitem{naik2011kurtosis}
G.~R. Naik, D.~K. Kumar, and S.~P. Arjunan, ``Kurtosis and negentropy
  investigation of myo electric signals during different {MVCs},'' in
  \emph{ISSNIP Biosignals and Biorobotics Conference 2011}.\hskip 1em plus
  0.5em minus 0.4em\relax IEEE, 2011, pp. 1--4.

\bibitem{zhao2012simulation}
Y.~Zhao and D.~Li, ``A simulation study on the relation between muscle motor
  unit numbers and the non-gaussianity/non-linearity levels of surface
  electromyography,'' \emph{Science China Life Sciences}, vol.~55, pp.
  958--967, 2012.

\bibitem{messaoudi2017assessment}
N.~Messaoudi, R.~E. Bekka, P.~Ravier, and R.~Harba, ``Assessment of the
  non-gaussianity and non-linearity levels of simulated semg signals on
  stationary segments,'' \emph{Journal of Electromyography and Kinesiology},
  vol.~32, pp. 70--82, 2017.

\bibitem{gini2000performance}
F.~Gini, M.~Greco, M.~Diani, and L.~Verrazzani, ``Performance analysis of two
  adaptive radar detectors against non-gaussian real sea clutter data,''
  \emph{IEEE Transactions on Aerospace and Electronic Systems}, vol.~36, no.~4,
  pp. 1429--1439, 2000.

\bibitem{yao2003spherically}
K.~Yao, ``Spherically invariant random processes: Theory and applications,''
  \emph{Communications, Information and Network Security}, pp. 315--331, 2003.

\bibitem{furui2021emg}
A.~Furui, T.~Igaue, and T.~Tsuji, ``{EMG} pattern recognition via {B}ayesian
  inference with scale mixture-based stochastic generative models,''
  \emph{Expert Systems with Applications}, vol. 185, p. 115644, 2021.

\bibitem{hayashi2017variance}
H.~Hayashi, A.~Furui, Y.~Kurita, and T.~Tsuji, ``A variance distribution model
  of surface emg signals based on inverse gamma distribution,'' \emph{IEEE
  Transactions on Biomedical Engineering}, vol.~64, no.~11, pp. 2672--2681,
  2017.

\bibitem{wang2006maximum}
J.~Wang, A.~Dogandzic, and A.~Nehorai, ``Maximum likelihood estimation of
  compound-gaussian clutter and target parameters,'' \emph{IEEE Transactions on
  Signal Processing}, vol.~54, no.~10, pp. 3884--3898, 2006.

\bibitem{richards2014fundamentals}
M.~A. Richards, \emph{Fundamentals of radar signal processing}.\hskip 1em plus
  0.5em minus 0.4em\relax McGraw-Hill Education, 2014.

\bibitem{gradshteyn2014table}
I.~S. Gradshteyn and I.~M. Ryzhik, \emph{Table of integrals, series, and
  products}.\hskip 1em plus 0.5em minus 0.4em\relax Academic press, 2014.

\bibitem{eltoft2006multivariate}
T.~Eltoft, T.~Kim, and T.-W. Lee, ``On the multivariate {L}aplace
  distribution,'' \emph{IEEE Signal Processing Letters}, vol.~13, no.~5, pp.
  300--303, 2006.

\bibitem{dempster1977maximum}
A.~P. Dempster, N.~M. Laird, and D.~B. Rubin, ``Maximum likelihood from
  incomplete data via the em algorithm,'' \emph{Journal of the royal
  statistical society: series B (methodological)}, vol.~39, no.~1, pp. 1--22,
  1977.

\bibitem{bishop2006pattern}
C.~M. Bishop and N.~M. Nasrabadi, \emph{Pattern recognition and machine
  learning}.\hskip 1em plus 0.5em minus 0.4em\relax Springer, 2006, vol.~4,
  no.~4.

\bibitem{8450847}
A.~C. Turlapaty, ``Shape parameter estimation for k-distribution using
  variational {B}ayesian approach,'' in \emph{2018 IEEE Statistical Signal
  Processing Workshop (SSP)}, 2018, pp. 243--247.

\bibitem{kelley2003solving}
C.~T. Kelley, \emph{Solving nonlinear equations with Newton's method}.\hskip
  1em plus 0.5em minus 0.4em\relax SIAM, 2003.

\bibitem{spanos2019probability}
A.~Spanos, \emph{Probability Theory and Statistical Inference: {E}mpirical
  Modeling with Observational {Data}}.\hskip 1em plus 0.5em minus 0.4em\relax
  Cambridge University Press, 2019.

\bibitem{kullback1997information}
S.~Kullback, \emph{Information theory and statistics}.\hskip 1em plus 0.5em
  minus 0.4em\relax Courier Corporation, 1997.

\bibitem{cohen1983applied}
J.~Cohen, P.~Cohen, S.~G. West, and L.~S. Aiken, ``Applied multiple
  regression,'' \emph{Correlation Analysis for the Behavioral Sciences},
  vol.~2, 1983.

\bibitem{pawitan2001all}
Y.~Pawitan, \emph{In all likelihood: statistical modelling and inference using
  likelihood}.\hskip 1em plus 0.5em minus 0.4em\relax Oxford University Press,
  2001.

\bibitem{santos2021classification}
E.~R.~T. Santos~Junior, B.~F. de~Salles, I.~Dias, A.~S. Ribeiro, R.~Sim{\~a}o,
  and J.~M. Willardson, ``Classification and determination model of resistance
  training status,'' \emph{Strength and Conditioning Journal}, vol.~43, no.~5,
  pp. 77--86, 2021.

\bibitem{mayhew1995muscular}
T.~P. Mayhew, J.~M. Rothstein, S.~D. Finucane, and R.~L. Lamb, ``Muscular
  adaptation to concentric and eccentric exercise at equal power levels,''
  \emph{Medicine and science in sports and exercise}, vol.~27, no.~6, pp.
  868--873, 1995.

\bibitem{baley1966effects}
J.~A. Baley, ``Effects of isometric exercises done with a belt upon the
  physical fitness status of students in required physical education classes,''
  \emph{Research Quarterly. American Association for Health, Physical Education
  and Recreation}, vol.~37, no.~3, pp. 291--301, 1966.

\bibitem{mardia1974applications}
K.~V. Mardia, ``Applications of some measures of multivariate skewness and
  kurtosis in testing normality and robustness studies,'' \emph{Sankhy{\=a}:
  The Indian Journal of Statistics, Series B}, pp. 115--128, 1974.

\bibitem{cain2017univariate}
M.~K. Cain, Z.~Zhang, and K.-H. Yuan, ``Univariate and multivariate skewness
  and kurtosis for measuring nonnormality: Prevalence, influence and
  estimation,'' \emph{Behavior research methods}, vol.~49, pp. 1716--1735,
  2017.

\bibitem{reed2008principles}
J.~Reed and J.~Bowen, ``Principles of sports rehabilitation,'' \emph{The sports
  medicine resource manual. 1st ed. Philadelphia: Saunders}, pp. 431--6, 2008.

\end{thebibliography}

\end{document}